%% file: main.tex
\documentclass[10pt,conference]{IEEEtran}

\IEEEoverridecommandlockouts
\input{ourcommands.tex}

\begin{document}

\title{
	Ellipsis: Towards Efficient System Auditing for Real-Time Systems
}

\author{
	\IEEEauthorblockN{
		     Ayoosh Bansal\IEEEauthorrefmark{1}\orcidicon{0000-0002-4848-6850},
		     Anant Kandikuppa\IEEEauthorrefmark{1},
		     Chien-Ying Chen\IEEEauthorrefmark{1},
		     Monowar Hasan\IEEEauthorrefmark{2}\orcidicon{0000-0002-2657-0402},
		     Adam Bates\IEEEauthorrefmark{1},
		     Sibin Mohan\IEEEauthorrefmark{3}\orcidicon{0000-0002-3295-0233}
	}
	
	\IEEEauthorblockA{
		     \IEEEauthorrefmark{1}
		     University of Illinois at Urbana-Champaign 
		     \{ayooshb2,
 		       anantk3,
		       cchen140,
		       batesa\}@illinois.edu
		\\
		     \IEEEauthorrefmark{2}
			 Wichita State University 
			 monowar.hasan@wichita.edu
		\\
			 \IEEEauthorrefmark{3}
			 Oregon State University 
			 sibin.mohan@oregonstate.edu
	}
}

\maketitle
\thispagestyle{plain}
\pagestyle{plain}

    \begin{abstract}
        \input{sections/abstract}

    \end{abstract}

    \input{sections/introduction.tex}

    \input{sections/threatmodel.tex}
    \input{sections/logreduction_new.tex}

    \input{sections/eval}
    \input{sections/security_analysis_new.tex}

    \input{sections/discussion.tex}

\input{sections/relatedwork.tex}

    \input{sections/conclusion.tex}
    \bibliography{ref_original,bates-bib-master}

\onecolumn
\twocolumn

    \appendices
    \newpage
    \input{sections/timing.tex}

    \newpage

\input{sections/appendix_tpl.tex}
    \newpage
    \input{sections/scaling_variability.tex}
    \newpage
    \input{sections/appendix_reconstruct.tex}

\end{document}

%% file: ourcommands.tex
\usepackage{xspace}
\usepackage{booktabs}
\usepackage{amsmath,amssymb,amsfonts}
\usepackage{algorithmic}
\usepackage{graphicx}
\usepackage{caption,subcaption}

\newcommand{\Sys}{{\textit{Ellipsis}}\xspace}
\newcommand{\SysBold}{{\textbf{Ellipsis}}\xspace}
\newcommand{\Systwo}{{\textit{Ellipsis-HP}}\xspace}
\newcommand{\SystwoBold}{{\textbf{Ellipsis-HP}}\xspace}

\usepackage{amssymb}

\usepackage{pifont}

\usepackage{listings}
\usepackage{lstautogobble}
\usepackage[dvipsnames]{xcolor}
\definecolor{codegreen}{rgb}{0,0.6,0}
\definecolor{codegray}{rgb}{0.5,0.5,0.5}
\definecolor{codepurple}{rgb}{0.58,0,0.82}
\definecolor{backcolour}{rgb}{0.95,0.95,0.92}
\definecolor{key-color}{rgb}{0.8, 0.47, 0.196}

\lstdefinestyle{mystyle}{
	commentstyle=\color{codegreen},
	keywordstyle=\color{magenta},
	numberstyle=\tiny\color{codegray},
	stringstyle=\color{codepurple},
	basicstyle=\ttfamily\scriptsize,
	breakatwhitespace=false,
	breaklines=true,
	captionpos=b,
	keepspaces=true,
	showspaces=false,
	showstringspaces=false,
	showtabs=false,
	tabsize=2,
    showlines=false,
    mathescape,
    morekeywords={$\varnothing$},
}
\usepackage{footmisc}

\usepackage{xkcdcolors} 
\usepackage{tikz}
\usepackage{wrapfig}
\usetikzlibrary{backgrounds}
\usetikzlibrary{arrows,shapes}
\usetikzlibrary{tikzmark}
\usetikzlibrary{calc}

\usepackage[edges]{forest}
\usetikzlibrary{arrows.meta}
\colorlet{linecol}{black!75}

\newcommand{\highlight}[2]{\colorbox{#1!17}{#2}}

\newcommand*\circled[1]{\tikz[baseline=(char.base)]{
            \node[shape=circle,draw,inner sep=1pt] (char) {#1};}}

\input{newcommands}

\newcommand\blfootnote[1]{%
	\begingroup
	\renewcommand\thefootnote{}\footnote{#1}%
	\addtocounter{footnote}{-1}%
	\endgroup
}

\newcommand{\inlinecode}[1]{%
	\texttt{%
		\setlength{\spaceskip}{0.5em plus 1em minus 0.1em}%
		\ifdim\lastskip>0pt \unskip\hspace{0.5em plus 0.5em minus 0.1em}\fi
		#1
	}%
}

\usepackage{scalerel}
\usepackage{tikz}
\usetikzlibrary{svg.path}
\definecolor{orcidlogocol}{HTML}{A6CE39}
\tikzset{
	orcidlogo/.pic={
		\fill[orcidlogocol] svg{M256,128c0,70.7-57.3,128-128,128C57.3,256,0,198.7,0,128C0,57.3,57.3,0,128,0C198.7,0,256,57.3,256,128z};
		\fill[white] svg{M86.3,186.2H70.9V79.1h15.4v48.4V186.2z}
		svg{M108.9,79.1h41.6c39.6,0,57,28.3,57,53.6c0,27.5-21.5,53.6-56.8,53.6h-41.8V79.1z M124.3,172.4h24.5c34.9,0,42.9-26.5,42.9-39.7c0-21.5-13.7-39.7-43.7-39.7h-23.7V172.4z}
		svg{M88.7,56.8c0,5.5-4.5,10.1-10.1,10.1c-5.6,0-10.1-4.6-10.1-10.1c0-5.6,4.5-10.1,10.1-10.1C84.2,46.7,88.7,51.3,88.7,56.8z};
	}
}
\newcommand\orcidicon[1]{\href{https://orcid.org/#1}{\mbox{\scalerel*{
				\begin{tikzpicture}[yscale=-1,transform shape]
					\pic{orcidlogo};
				\end{tikzpicture}
			}{|}}}}

\usepackage[colorlinks,bookmarks=true,hypertexnames=true]{hyperref}


%% file: newcommands.tex
\newcommand{\eg}{{\it e.g.,}\xspace}

\newcommand{\etal}{{\it et~al.}\xspace}
\newcommand{\ie}{{\it i.e.,}\xspace}
\newcommand{\etc}{{\it etc.~}\xspace}
\newcommand{\ci}{{\it (i) }}
\newcommand{\cii}{{\it (ii) }}

\newcommand{\ca}{{\it (a) }}
\newcommand{\cb}{{\it (b) }}
\newcommand{\cc}{{\it (c) }}



%% file: sections/abstract.tex
System auditing is a powerful tool that provides insight
  into the nature of suspicious events in computing systems,
  allowing machine operators to detect and subsequently investigate
  security incidents.
While auditing has proven invaluable to the security of traditional
  computers, existing audit frameworks are rarely designed with consideration
  for Real-Time Systems (RTS).
The transparency provided by system auditing would be of
  tremendous benefit in a variety of security-critical RTS domains,
  (\eg autonomous vehicles);
  however, if audit mechanisms are not carefully integrated into RTS,
  auditing can be rendered ineffectual and violate the real-world temporal requirements of the RTS.

In this paper, we demonstrate how to adapt commodity audit frameworks to RTS.
Using Linux Audit as a case study,
  we first demonstrate that the volume of audit events generated by commodity frameworks
  is unsustainable within the temporal and resource constraints
  of real-time (RT) applications.
To address this, we present \Sys,
  a set of kernel-based reduction techniques that leverage the periodic
  repetitive nature of RT  applications to aggressively reduce the
  costs of system-level auditing.
\Sys generates succinct descriptions of RT applications' expected activity
  while retaining a detailed record of unexpected activities,
  enabling analysis of suspicious activity while
  meeting temporal constraints.
Our evaluation of \Sys, using ArduPilot (an open-source autopilot application suite)
  demonstrates \textit{up to 93\% reduction} in audit log generation.
\Sys demonstrates a promising path forward for auditing RTS.

%% file: sections/introduction.tex
\section{Introduction}
\label{sec:introduction}

\blfootnote{
	This preprint has not undergone peer review or any post-submission improvements or corrections.
}

As RTS become indispensable in safety- and security-critical domains ---
  medical devices, autonomous vehicles, manufacturing automation,
  smart cities, etc.
  \cite{gurgen2013self,lee2011challenges,monostori2016cyber,rajkumar2010cyber} ---
  the need for effective and precise \textit{auditing} support is growing.
Even now, event data recorders (or \textit{black boxes}) are crucial for determining fault and liability
  when investigating vehicle collisions  \cite{edr2,edr1, edr_blame1,edr_blame2}
  and the need for diagnostic event logging frameworks (\eg
  QNX~\cite{qnx_logging}, VxWorks~\cite{vxworks_logging} and Composite OS~\cite{parmer2010composite,song2015c})  is  well understood.
However, these high-level event loggers are insufficient to detect
  and investigate sophisticated attacks.
Concomitant with its explosive growth, today's RTS have become ripe targets for sophisticated attackers~\cite{dhs_cpssec}.
Exploits in RTS can enable vehicle hijacks~\cite{cps_article_7,cps_article_6},
  manufacturing disruptions~\cite{cps_article_8}, IoT botnets~\cite{cps_article_5}, subversion of life-saving medical devices~\cite{cps_article_4} and many other devastating attacks.
The COVID-19 pandemic has further shed light on the potential damage of attacks on medical infrastructure~\cite{cps_article_2,cps_article_3}.
These threats are not theoretical, rather active and ongoing, as evidenced recently by malicious attempts to take control of nuclear power, water and electric systems throughout the United States and Europe~\cite{ps2018_nyt}.

In traditional computing systems, {\it system auditing} has proven crucial to detecting, investigating and responding to intrusions~\cite{carbonblackreport,hgl+2019,hnd+2020,mge+2019}.
System auditing takes place at the \textit{kernel layer} and creates a new event for every syscall that is issued.
Not only does this approach take the responsibility of event logging out of the hands of the application developer, it also provides a unified view of system activity in a way that application-specific logging simply cannot.
In particular, systems logs can be iteratively parsed into a connected graph based on the shared dependencies of individual events, facilitating causal analysis over the history of events within a system \cite{kc2003,btb+2015,pgs+2016,mzx2016,hmw+2017,phg+2017,kww+2018,kww+2018,lzl+2018,meg+2018,tzl+2018}.
This capability is invaluable to defenders when tracing suspicious activities \cite{hgl+2019,mge+2019,hnd+2020}, to the point that the vast majority of cyber analysts consider audit logs to be the most important resource when investigating threats~\cite{carbonblackreport}.
Hence, the deployment of system-level audit capabilities can help on multiple fronts:
  \ca fault detection/diagnosis and
  \cb understanding and detecting security events.

Unfortunately, comprehensive system auditing approaches are not
  widely used in RTS.
RTS logging takes place largely at the {\it application layer}~\cite{edr2,edr1}
  or performs lightweight system layer tracing for performance profiling
   (\eg log syscall occurrences, but not arguments)~\cite{brandenburg2007feather};
  in both cases, the information recorded is
  insufficient to trace attacks because the causal links between different system entities cannot be identified.
The likely cause of this hesitance to embrace holistic system-layer logging is poor performance. System audit frameworks are known to impose tremendous computational and storage overheads~\cite{mzk+2018} that are incompatible with the temporal requirements of many real-time applications.
Furthermore, system auditing can introduce unpredictable behaviors (say, due to the need to flush out a full audit buffer)
and timing perturbations, not to mention priority inversions and inter-application contentions --- all of which can have significant negative impacts on safety.
Section~\ref{sec:basetiming} provides a detailed discussion of these potential issues.
Thus, while we are encouraged by the growing recognition of the importance of embedded system auditing
\cite{embedded_audit_2,embedded_audit_3,embedded_audit_1}
and the newfound availability of Linux Audit in the Embedded Linux  \cite{elinux}, a practical approach to RTS auditing remains an elusive goal.

\begin{figure}[t]
	\centering
	\includegraphics[width=\linewidth]{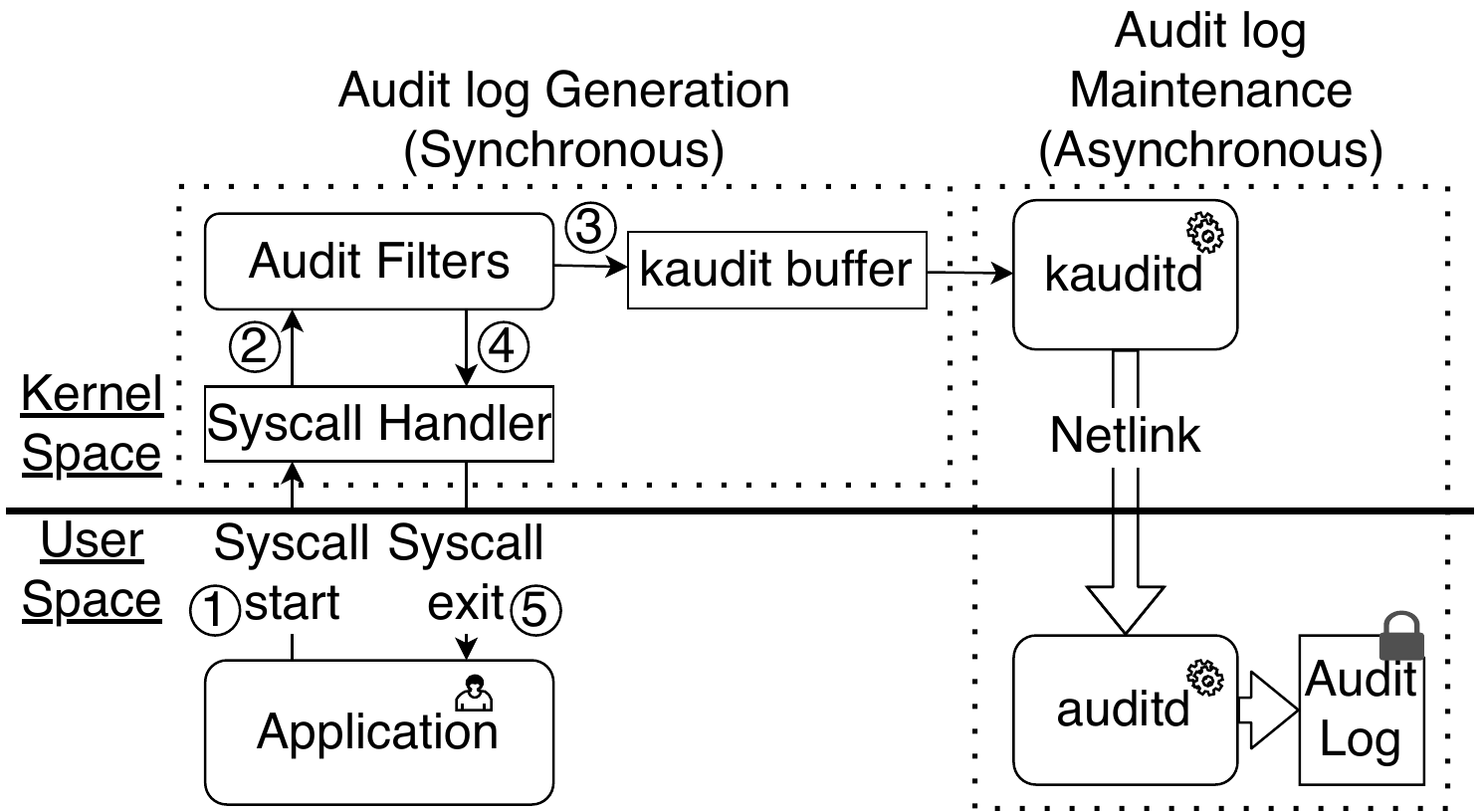}
	\caption{\label{fig:audit_arch}
		Architecture of Linux Audit Framework~\cite{audit}.
	}
\end{figure}

This work presents a thorough exploration of aggressive system auditing within
  RT environments.
We first conduct an analysis of Linux Audit's adherence to real-time scheduling principles,
  discovering that while Audit introduces overheads and increased variance to each syscall,
  it does not introduce inter-application resource contention or priority inversion~($\S$\ref{sec:basetiming}).
Observing that performance cost of Linux Audit is ultimately dependent
  on the number of log events generated,
and that the performance impacts of commodity auditing frameworks can be optimized without affecting the forensic validity of the audit logs,\textit{ e.g.}, through carefully reducing the number of events that need to be logged \cite{bbm2015,bth+2017,bhc+2018,hal+2018,lzx2013,mlk+2015,mzk+2018,tll+2018,xwl+2016},
  we set out to tailor Linux Audit to RTS,
carefully reducing event logging without impacting the forensic validity of the log.
We present \mbox{\SysBold}\blfootnote{\label{foot:source}\url{https://bitbucket.org/sts-lab/ellipsis}}, a kernel-based log reduction framework that leverages the
  predictability of real-time tasksets' execution profiles.
\Sys first profiles tasks to produce a \textit{template} of their audit footprint.
At runtime, behaviors consistent with this template are reduced,
  while any deviations from the template are audited in full, without reduction.
Far from being impractical, we demonstrate a synergistic relationship between security auditing and predictable RTS workloads -- \Sys is able to
  faithfully audit suspicious activities while incurring almost no
  log generation during benign typical activity.
%
%
The \textbf{contributions} of this work are:
\begin{itemize}
    \item \Sys, an audit framework, uniquely-tailored to RT environments ($\S$\ref{sec:logreduction}).
     To the best of our knowledge, \Sys is the first auditing framework designed to exploit the well-formed nature of real-time tasks.

    \item Detailed evaluations ($\S$\ref{sec:eval}) and security analysis ($\S$\ref{sec:security_analysis}) to demonstrate that \Sys retains relevant information while reducing audit event and log volume.
\end{itemize}

%% file: sections/threatmodel.tex
\section{Background and System Model}
\label{sec:threatmodel}

\subsection{Linux Audit Framework}

The Linux Audit system \cite{suse2004} provides a way to audit system activities.
An overview of the Linux Audit architecture is presented in Figure \ref{fig:audit_arch}.
When an application invokes a syscall \circled{1},
the subsequent kernel control flow eventually traverses an {\tt audit\_filter} hook \circled{2}.
Linux Audit examines the context of the event, compares it to
pre-configured audit rules,
generating a log event if there is a match and enqueueing it
in a  message buffer  \circled{3}
before returning control to the syscall handler \circled{4} and then to the application \circled{5}.
Asynchronous from this workflow, a pair of (non-RT) audit daemons
({\tt kauditd} and {\tt auditd})
work in tandem to
transmit the message buffer
to user space for storage and analysis.
Because the daemons are asynchronous,
the message buffer can overflow if
syscalls occur faster than the daemon flushes audit records to user space,
resulting in event loss.

Although it is well-established that Linux Audit can incur large computational and storage overheads in traditional software~\cite{mzk+2018}, its impacts on RT applications were unclear.
Linux Audit not only adds additional latency to each syscall as it generates log messages
but also introduces the shared {\tt kauditd} buffer whose access is coordinated using a spinlock.
These changes could potentially wreak havoc on RT task sets as a result of
changing execution profiles, resource contention or priority inversion~\cite{priority_inversion}.
Encouragingly, upon conducting a detailed analysis (Appendix~\ref{sec:basetiming})
  we observed that Linux Audit does not introduce significant issues of priority inversion or contention over auditing resources shared across applications (\eg \xspace {\tt kaudit} buffer).
Further, except for limited outlier cases, the latency introduced by auditing syscalls can be measured and bounded.
Hence it is a good candidate for firm and soft deadline RTS as supported by RT Linux~\cite{preemptrt}.
However, audit events were lost, making auditing incomplete and ineffectual while still costly for the RTS due to large storage space required to store the audit log.
While Linux Audit can be configured to monitor high-level activities such as login attempts \cite{audit}, its primary utility (and overhead) comes from tracking low-level syscalls, which is the focus of this paper.

\begin{table*}[t]
	\centering
	\caption{\label{tab:rts_properties}
		RTS properties relevant to \Sys}
		\begin{tabular}{p{0.1\linewidth} | p{0.65\linewidth} | p{0.15\linewidth}
			}
			\toprule
			Property & Relevance to \Sys & Sections
			\\
			\midrule
			
			Periodic tasks
			& Most RT tasks are periodically activated, leading to repeating behaviors. \Sys templates describe the most common repetitions.
			& \ref{sec:repeated_sequences}, \ref{sub:template}
			\\
			\midrule
			
			Aperiodic tasks
			& Second most common form of RT tasks, Aperiodic tasks also lead to repeating behaviors, but with irregular inter-arrival times.
			& \ref{sec:repeated_sequences}, \ref{sub:template}
			\\
			\midrule
			
			Code Coverage
			& High code coverage analyses are part of existing RTS development processes, \Sys' automated template generation adds minimal cost.
			& \ref{sec:code_coverage}, \ref{sec:discuss_code_coverage}
			\\
			\midrule
			
			Timing Predictability
			& A requirement for safety and correct functioning of RTS, na{\"i}vely enabling auditing can violate this by introducing overheads and variability.
			& \ref{sec:temporal_contraints}, \ref{sec:discussion}, \ref{sec:temporal_1}
			\\
			\midrule

			Isolation
			& Resources are commonly isolated in RTS to improve timing predictability. RTS auditing mechanisms should not violate resource isolation.
			& \ref{sub:priority_inversion}, \ref{sub:contention}
			\\
			\midrule

			Special Purpose
			& RTS are special purpose machines, tasks are known at development \ie templates can be created before system deployment.
			& \ref{sub:template}, \ref{sec:discussion}
			\\
			\midrule
			
			Longevity
			& Once deployed RTS can remain functional for years. \Sys' can save enormous log storage and transmission costs over the lifetime of the RTS.
			& \ref{sec:eval_log_reduction}
			\\
			
			\bottomrule
		\end{tabular}
\end{table*}

\subsection{RTS Properties}
\label{sec:rts_concepts}
\Sys leverages properties unique to RT environments, 
that we describe here.
In contrast to traditional applications where determining all possible
execution paths is often undecidable,
knowledge about execution paths is an essential component of RT application development.
RTS are special purpose machines that execute well formed tasksets to fulfill predetermined tasks.
RT Applications structure commonly involve repeating loops that are excellent targets for conversion to templates.
Various techniques are employed to analyze the tasksets with high code coverage
\eg worst case execution time (WCET) analysis for real time tasks
\cite{burguiere2006history,gustafsson2007experiences,hatton2004safer,Liu:1973:SAM:321738.321743,puschner2002writing,sandell2004static,yoon2017learning}.
All expected behaviors of the system must be accounted for at design time in conjunction with the system designers.
Any deviation is an unforeseen fault or malicious activity, which needs to be audited in full detail.
Table~\ref{tab:rts_properties} contains a summary of RTS features and constraints,
with references to sections in this work that discuss, evaluate or leverage these features.
In this work we show that applications from two different classes of RTS follow this model;
a control application (Ardupilot~\cite{ardupilot}) and a video analysis application (Motion~\cite{motion}).

\subsubsection{Repetition of Sequences}
\label{sec:repeated_sequences}
In their seminal work on Intrusion Detection, Hofmeyr~\etal~\cite{hofmeyr1998intrusion} established that normal behavior of an application can be profiled as sequences of syscall.
This works exceptionally well for RTS as they feature limited tasks with limited execution paths on a system.
Unlike general purpose systems, RTS run limited predefined tasks.
The requirements of reliability, safety and timing predictability imply that RTS have limited execution paths which can be tested and analyzed to ensure the previously mentioned requirements.
The RTS survey~\cite{akesson2020empirical} further found that 82\% of the RTS contained tasks with periodic activation.
Periodically activated tasks with limited execution paths will invariably lead to high repetitions of certain sequences.
Yoon~\etal~\cite{yoon2017learning} demonstrated the existence of repeating syscall sequences in a RTS.
The reliable repetition of behaviors has also led to profile driven techniques being successfully employed towards achieving predictable temporal behaviors in RTS~\cite{BBProf_ECRTS21}.

\subsubsection{Code Coverage}
\label{sec:code_coverage}
A recently published survey of industry practitioners in RTS~\cite{akesson2020empirical} noted that the five most important system aspects for industrial RTS were
Functional correctness, Reliability and availability, System safety, Timing predictability, System security.
The role of code coverage in software testing is well established~\cite{wong2007effective,ivankovic2019code}.
Prior works have established the correlation between code coverage and reliability of software~\cite{del1995correlation,chen1996empirical,chen2001effect}.
Software safety standards include structural code coverage as a requirement~\cite{rtca1992software,bordin2009couverture}.
Timing predictability in RTS is ensured by coding standards/guidelines~\cite{misra,hatton2004safer,puschner2002writing} and worst case execution time (WCET) analysis~\cite{sandell2004static,gustafsson2007experiences}.
Therefore, high code coverage is an integral component of RTS development process.
Template generation for \Sys therefore does not introduce a significant additional burden.
Template sequences are determinable in the course of existing development processes for RTS.

\subsection{Threat Model}
%
We consider an adversary that aims to penetrate and impact an RTS through
  exfiltration of data, corruption of actuation outputs, degradation of performance,
  causing deadline violations, \etc.
This attacker may install modified programs, exploit a running process or install malware on the RTS to achieve their objectives.
To observe this attacker, our system adopts
  an aggressive audit configuration intended to capture all forensically-relevant events, as identified in prior works.%
\footnote{\label{foot:audit_rules} Specifically, our ruleset audits \inlinecode{execve, read, readv, write, writev, sendto, recvfrom, sendmsg, recvmsg, mmap, mprotect, link, symlink, clone, fork, vfork, open, close, creat, openat, mknodat, mknod, dup, dup2, dup3, bind, accept, accept4, connect, rename, setuid, setreuid, setresuid, chmod, fchmod, pipe, pipe2, truncate, ftruncate, sendfile, unlink,  unlinkat, socketpair,splice, init\_module,} and \inlinecode{finit\_module.}}
  We assume that the underlying OS and the audit subsystem therein are trusted.
  This is a standard assumption in system auditing literature
  \cite{btb+2015,hgl+2019,Liu2018TowardsAT,Ma2016ProTracerTP,pmm+2012}.
  Far from being impractical on RTS, prior works such as Trusted Timely Computing Base provide a secure kernel that meets both the trust and temporal requirements for hosting \Sys in RT Linux~\cite{casimiro2000build,correia2002design,verissimo2002timely,verissimo2000timely}.

\Sys' goal is to capture evidence of an attacker intrusion/activity without losing relevant information and hand it off to a tamper proof system.
Although audit integrity is an important security goal,
  it is commonly explored orthogonally to other audit research due to the
  modularity of security solutions (\eg \cite{btb+2015,pdh+2020,ynh+2021}).
  Therefore, we assume that once recorded to {\tt kaudit buffer}, attackers cannot compromise the integrity of audit logs
Finally, we assume that applications can be profiled in a controlled benign environment
prior to being the target of attack, such as pre-deployment testing and verification.



%% file: sections/logreduction_new.tex
\section{\Sys}
\label{sec:logreduction}
\label{sub:template}

The volume of audit events is the major limiting factor for auditing RTS.
High event volume can result in event record loss, high log storage costs and large maintenance overheads~\cite{mzk+2018}.
We present \Sys, an audit event reduction technique designed specifically for RTS.
\Sys achieves this through {\it templatization} of the audit event stream.
Templates represent learned expected behaviors of RT tasks,
described as a sequence of syscalls with arguments and temporal profile%
\footnote{\label{foot:templates}
	 Template examples are available as Appendix~\ref{sec:appendix_templates}}.
These templates are generated in an offline profiling phase, similar to common RTS analyses like WCET~\cite{burguiere2006history,li1995performance}.
At runtime, the application's syscall stream is compared against its templates;
if a contiguous sequence of syscalls matches a template,
only a single record indicating the template match is inserted into the event stream ({\tt kaudit buffer}).
Matched syscalls are never inserted into the event stream, reducing the number of events generated by the auditing system.
Significantly, while a sequence of audited syscall events is replaced by a single record, relevant information is not lost ($\S$\ref{sec:security_analysis}).

 \begin{table}[t]%
	 \centering%
	 \caption{Symbols Summary}%
	 \label{tab:symbols}%
	 \begin{tabular}{cl}%
		 \toprule
		 \textbf{Symbol} & \textbf{Description}
		\\ \midrule
		 S1, S2, S3 & System Calls
		 \\ \midrule
		 $TPL-X$ & Template-X where $X \in \mathbb{N}$
		 \\ \midrule
		 $\Delta_i$ & Observed runtime for ith instance of templatized task
		 \\ \midrule
		 $T_X$ & Temporal constraint of $TPL-X$
		 \\ \midrule
		 x,\{y\} & Unique state ID for \Sys FSA \\ & x: number of system calls matched \\ & y: set of potential template matches.
		 \\ \midrule
		 $\tau$ & A RT task
		 \\ \midrule
		 $s_i$ & $i^{}th$ syscall sequence exhibited by $\tau$
		 \\ \midrule
		 $N$ & Count of possible $s_i$ for $\tau$, $0 < i \leq N$
		 \\ \midrule
		 $len(s_i)$ & number of syscalls in $s_i$
		 \\ \midrule
		 $p_i$ & probability of occurence of corresponding $s_i$
		 \\ \bottomrule
		 \end{tabular}%
	 \end{table}%

\subsection{Model}
%
Consider a system in which the machine operator wishes to
audit a single RT task {\boldmath$\tau$}.
An RT \textit{task} here corresponds to a \textit{thread} in Linux systems, identified by a combination of process and thread ids.
We can limit this discussion to a single task, without losing generality, as \Sys' template creation, activation and runtime matching treat each task as independent.
Furthermore, we modified Linux Audit to include thread ids in audit event records.

RT tasks are commonly structured with a one time {\it init} component
and repeating {\it loops}.
Let {\boldmath$s_i$} denote a syscall sequence the task exhibits in a {\it loop} execution and {\boldmath$N$} the count of different syscall execution paths $\tau$ might take \mbox{(\ie $0 < i \leq N$)}.
A \textit{template} describes these sequences ($s_i$), identifying the syscalls and arguments.
As noted in Section~\ref{sec:rts_concepts}, RT applications are developed to have limited code paths and bounded loop iterations.
Extensive analysis of execution paths is a standard part of the RTS development process.
Thus, for RTS, $N$ is finite and determinable.
Let function {\boldmath$len(s_i)$} return the number of syscalls in the sequence $s_i$.
Further, let {\boldmath$p_i$} be the probability that an iteration of $\tau$ exhibits syscall sequence $s_i$.

\begin{figure}[t]
	\centering
	\includegraphics[width=.7\linewidth,keepaspectratio]{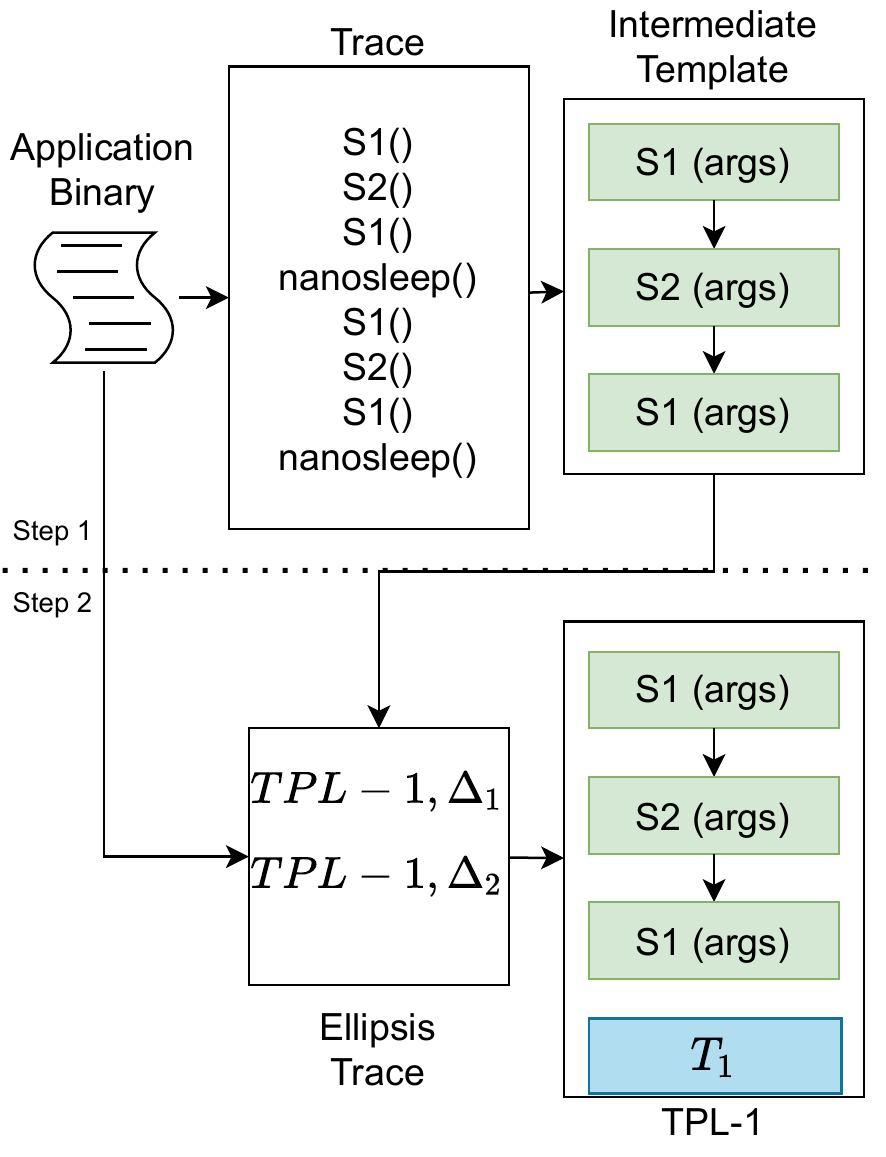}
	\caption{\label{fig:template}
		\Sys template creation.
        Syscalls are denoted by S1 and S2.
        Application is traced to identify repeating syscall sequences,
        then audited with \Sys, using the intermediate template, to get temporal constraint ($T_1$).
        This creates a profile for task's execution time ($\Delta_i$), yielding temporal constraints ($T_1$).
        Intermediate templates enriched with temporal constraints are the final templates.
	}
\end{figure}

\subsection{Sequence Identification}
The first step towards template creation is identification of sequences and their probability of occurrences.
Identification of cyclic syscall behaviors has been addressed in the auditing literature~\cite{lzx2013-beep,mzw+2017}, with past solutions require binary analysis, code annotations, stack analysis or a combination.
While any technique that yields $s_i$ and $p_i$ can be employed here, including the prior mentioned ones, we developed a highly automated process, leveraging RT task structure and Linux Audit itself.
The application is run for long periods of time and audit trace collected.
We observe that RT tasks typically end with calls to {\tt sleep} or {\tt yield} that translate to {\tt nanosleep} and {\tt sched\_yield} syscalls in Linux. Periodic behaviors can also be triggered by polling timerfds to read events from multiple timers by using {\tt select} and {\tt epoll\_wait} syscalls.
We leverage these syscalls to identify boundaries of task executions within the audit log and then extract sequences of syscall invocations.
Figure~\ref{fig:template} provides an overview of this process.
We also modified Linux Audit to include the Thread ID in log messages helping disambiguate threads belonging to a process, yielding thread level sequences.
This first step yields the per task syscall sequences exhibited by the application and their properties: length, probability of occurrence, and the arguments.
These syscall sequences are then converted into intermediate thread-level templates, each entry of which includes the syscall name along with the arguments.
This first step can also be iterated with intermediate templates loaded to reduce previously extracted sequences, though in practice such iterations were not required.

\subsection{Sequence Selection}
A subset of intermediate templates are chosen to be converted to final templates.
This choice is based on the tradeoff between the benefit of audit event volume reduction and the memory cost as defined later in eq. \eqref{eq:sys_save_event} and \eqref{eq:runtime_mem}, respectively.
As we discuss in detail in Section~\ref{sec:security_analysis} the security tradeoff is minimal.
Let's assume {\boldmath$n$} sequences are chosen to be reduced, where $0 \leq n \leq N$.
As noted earlier, \Sys treats each task independently, the value of $n$ is also independent for each task.

\subsection{Template Creation}
\label{sec:template_step2}
For the next step, Figure~\ref{fig:template} Step 2, these $n$ templates are loaded and application profiled again to collect temporal profile for each template \ie the expected duration and inter-arrival intervals for each template.
The intermediate templates are enriched with this temporal information, to yield the final templates.
Templates are stored in the form of text files and occupy negligible disk space,
\eg ArduPilot templates used for evaluation ($\S$\ref{sec:eval}) occupied 494 bytes of space on disk total.
This whole process is highly automated, given an application binary with necessary inputs, using the template creation toolset.\footref{foot:source}

\subsection{\Sys Activation}
We extend the Linux Audit command-line {\tt auditctl} utility to transmit templates to kernel space.
Once templates are loaded, \Sys can be activated using {\tt auditctl} to start reducing any matching behaviors.
This extended \texttt{auditctl} can also be used to activate/deactivate \Sys and load/unload templates,
however, these operations are privileged, identical to deactivating Linux Audit itself.
System administrators can use this utility to easily update templates as required, \eg in response to application updates.

\begin{figure}[t]
		\centering
		\includegraphics[width=.9\linewidth,keepaspectratio]{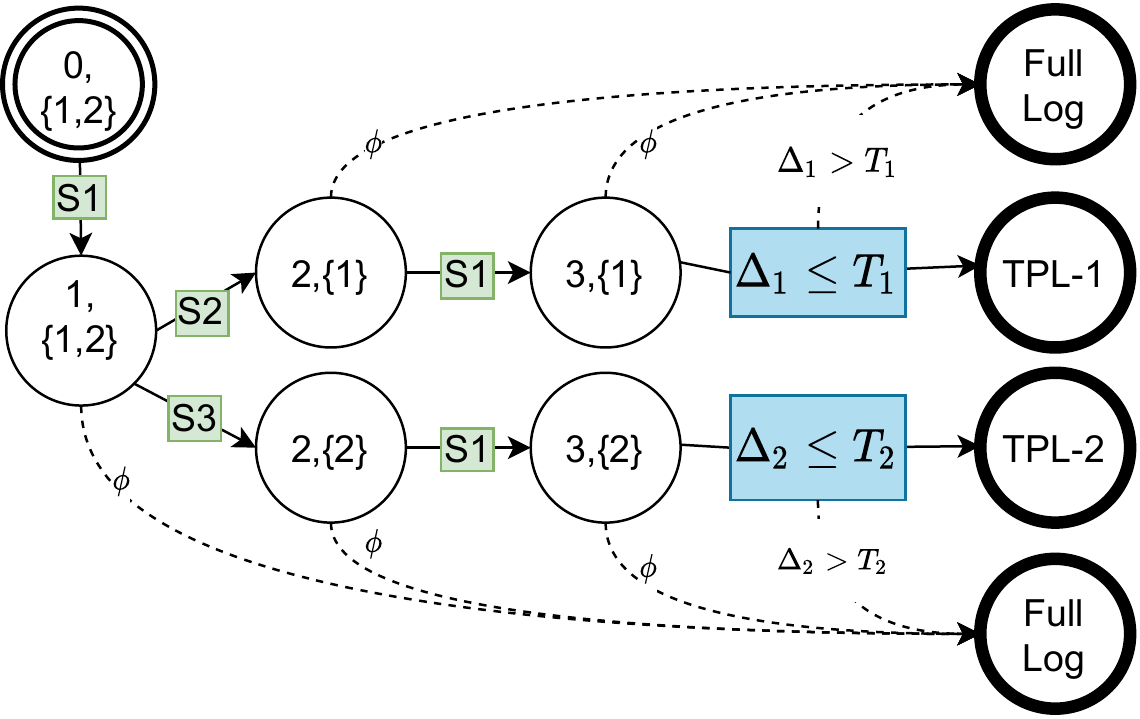}
		\caption{\label{fig:templ_prefix_tree}
			Runtime template matching as an FSA with states as [syscalls matched count, \{set of reachable templates\}].
            Syscall invocations trigger state transitions.			
			TPL-1 (S1, S2, S1) and TPL-2 (S1, S3, S1) are shown as example.
			Template matches (TPL-1, TPL-2) emit a single record, failure leads to full log store.
            Completing the template sequence and satisfying temporal constraints leads to an accept state (TPL-1,TPL-2) emitting a single record.
            Any divergence or failure causes \Sys to emit complete logs, shown as dotted transitions.
            The FSA then returns to initial state for the task to start capturing the next iteration.
		}
\end{figure}

\subsection{Runtime Matching}
\label{sec:template_runtime_match}
\label{sub:matching}
\label{sub:temp_tradeoff}
Given the template(s) of syscall sequences,
an \Sys kernel module, extending from Linux Audit syscall hooks,
filters syscalls that match a template.
The templates are modeled as a finite state automaton (FSA),
(Figure~\ref{fig:templ_prefix_tree}),
implemented as a collection of linked lists in kernel memory.
While the RT task is executing,
all syscall sequences allowed by the automaton are stored in a temporary task-specific buffer.
If the set of events fully describes an automaton template,
\Sys discards the contents of the task-specific buffer and enqueues a single record onto the {\tt kaudit} buffer to denote the execution of a
templatized activity.
Alternatively, \Sys enqueues the entire task-specific buffer to the main kaudit buffer if \ca a syscall occurs that is not allowed by the automaton, \cb the template is not fully described at the end of the task instance or \cc  the task instance does not adhere to the expected temporal behavior of the fully described template.
Thus, the behavior of each task instance is reduced to a single record
when the task behaves as expected.
For any abnormal behavior, the complete audit log is retained.

\begin{table*}[t]
	\centering
	\caption{Parameters from Case Study}%
	\label{tab:ardu_eq}%
			\begin{tabular}{lccccc}
				\toprule
				Task Name   & $N$ & $I$ & $len(s_i)$ & $p_i$ & $f$ \\
				\midrule
				arducopter  & 5  & 100  &   $[14,15,17,17,18]$        & $[0.95,0.02,0.01,0.01,0.01]$    & 679\\

				ap-rcin     & 1  & 182  &   $[16]$        & $[1$]    & 2 \\

				ap-spi-0    & 5  & 1599  & $[1,1,1,2,2]$          & $[0.645, 0.182, 0.170, 0.001,0.001]$    & 0  \\
				\bottomrule
			\end{tabular}
\end{table*}

\subsection{Audit Event Reduction}
Let the task $\tau$ be executed for {\boldmath$I$} iterations
and {\boldmath$f$} denote the number of audit events in {\it init} phase.
The number of audit events generated by {\boldmath$\tau$} when audited by Linux Audit {\boldmath$(E_A)$}, when \Sys reduces $n$ out of total $N$ sequences {\boldmath$(E_E)$}, and the reduction ($E_A - E_E$) are given by
\begin{align}
	E_A &=  I * (\textstyle \sum_{i=1}^{N}(p_i * len(s_i))) + f  \label{eq:audit_events} \\
	E_E &= I * (\textstyle \sum_{i=1}^{n}p_i +  \textstyle  \sum_{i=n+1}^{N}(p_i * len(s_i)))  + f  \label{eq:ellipsis_events} \\
	%
	%
	\nonumber \\
	\tikzmarknode{a}{\highlight{Bittersweet}{$E_A - E_E$}} &=
	\tikzmarknode{b}{\highlight{NavyBlue}{ $I$ }} * (
	\tikzmarknode{d}{\highlight{Fuchsia}{$\sum_{i=1}^{n}(p_i * len(s_i)$)}} -
	\tikzmarknode{c}{\highlight{OliveGreen}{$\sum_{i=1}^{n}p_i$}}) \label{eq:sys_save_event}
\end{align}
\begin{tikzpicture}[overlay,remember picture,>=stealth,nodes={align=left,inner ysep=1pt},<-]
	\path (a.north) ++ (.01,.6em) node[anchor=south west,color=Bittersweet!85] (mitext){\textsf{\footnotesize \Sys' Event reduction}};
	\draw [color=Bittersweet!85](a.north) |- ([xshift=-0.3ex,color=Bittersweet]mitext.south east);
	\path (b.south) ++ (0,-1.3em) node[anchor=south east,color=NavyBlue!85] (mitext){\textsf{\footnotesize Iterations}};
	\draw [color=NavyBlue!85](b.south) |- ([xshift=-0.3ex,color=NavyBlue]mitext.south west);
	\path (d.north) ++ (0,.5em) node[anchor=south west,color=Fuchsia!85] (mitext){\textsf{\footnotesize Audit events for n sequences}};
	\draw [color=Fuchsia!85](d.north) |- ([xshift=-0.3ex,color=Fuchsia]mitext.south east);
	\path (c.south) ++ (0,-.01em) node[anchor=north east,color=OliveGreen!85] (mitext){\textsf{\footnotesize \Sys events for n sequences}};
	\draw [color=OliveGreen!85](c.south) |- ([xshift=-0.3ex,color=OliveGreen!85]mitext.south west);
\end{tikzpicture}

As evident from eq.~\eqref{eq:sys_save_event}, to maximize reduction, long sequences with large $p_i$ values must be chosen as the $n$ sequences for reduction.
RT applications, like control systems, autonomous systems and even video streaming, feature limited execution paths for majority of their runtimes~\cite{konrad2005real}.
This property has been utilized by Yoon~\etal in a prior work~\cite{yoon2017learning}.
Therefore, for RT applications the distribution of $p_i$ is highly biased \ie certain sequences $s_i$ have high probability of occurrence.
Table~\ref{tab:ardu_eq} provides example values for the parameters used,
determined during the \textit{Sequence Identification} step in template creation for the evaluation application ArduPilot($\S$\ref{sec:eval}).

\subsection{Storage Size Reduction}
Let {\boldmath$B_A$} denote the average cost of representing a syscall event in audit log and {\boldmath$B_E$} denote the average cost of representing \Sys' template match record.
Thus $B_A$ represents the average size over all events in the Linux Audit log, whereas in \Sys syscall sequences that match a template will be removed and replaced with a template match event of an average size $B_E$.
By design, $B_E <= B_A$; $B_E$ is a constant 343 bytes, while $B_A$ averaged 527 bytes (1220 bytes max) in our evaluation.
Noting that the init events ($f$) are not reduced by \Sys, the disk size reduction \ie difference in sizes of {\boldmath$\tau$}'s audit log for Linux Audit {\boldmath$(L_A)$} and \Sys\ {\boldmath$(L_E)$} is:


\begin{align}
    L_A &=  I * (B_A * \textstyle \sum_{i=1}^{N}(p_i * len(s_i))) + f * B_A \label{eq:audit_log_size}
\end{align}
\begin{multline}
    L_E = I * (B_E * \textstyle \sum_{i=1}^{n}p_i + B_A * \textstyle  \sum_{i=n+1}^{N}(p_i * len(s_i)))  + \\ f * B_A \label{eq:ellipsis_log_size}
\end{multline}

The reduction in log size is given by:

\begin{multline}
	\tikzmarknode{a}{\highlight{Bittersweet}{$L_A - L_E$}} =
	\tikzmarknode{b}{\highlight{NavyBlue}{I}} * (
	\tikzmarknode{d}{\highlight{Fuchsia}{$B_A * \sum_{i=1}^{n}(p_i * len(s_i)$)}} -  \\
	\tikzmarknode{c}{\highlight{OliveGreen}{$B_E * \textstyle \sum_{i=1}^{n}p_i$}}) \label{eq:sys_save}
\end{multline}
\begin{tikzpicture}[overlay,remember picture,>=stealth,nodes={align=left,inner ysep=1pt},<-]
	\path (a.north) ++ (.1,.5em) node[anchor=south west,color=Bittersweet!85] (mitext){\textsf{\footnotesize Log size reduction}};
	\draw [color=Bittersweet!85](a.north) |- ([xshift=-0.3ex,color=Bittersweet]mitext.south east);
	\path (b.south) ++ (0,-1.3em) node[anchor=south east,color=NavyBlue!85] (mitext){\textsf{\footnotesize Iterations}};
	\draw [color=NavyBlue!85](b.south) |- ([xshift=-0.3ex,color=NavyBlue]mitext.south west);
	\path (d.north) ++ (-.2,.4em) node[anchor=south west,color=Fuchsia!85] (mitext){\textsf{\footnotesize Audit log size for n sequences}};
	\draw [color=Fuchsia!85](d.north) |- ([xshift=-0.3ex,color=Fuchsia]mitext.south east);
	\path (c.south) ++ (0,-.01em) node[anchor=north east,color=OliveGreen!85] (mitext){\textsf{\footnotesize \Sys log size for n sequences}};
	\draw [color=OliveGreen!85](c.south) |- ([xshift=-0.3ex,color=OliveGreen!85]mitext.south west);
\end{tikzpicture}

From \eqref{eq:sys_save_event}~and~\eqref{eq:sys_save},
\Sys' benefits come from the audit events count and log size becoming independent
of sequence size ($len(s_i)$) for the chosen $n$ sequences, multiplied further by repetitions of these sequences ($I * p_i$).
\Sys behaves identical to Linux Audit for any sequence that is not included as a template, \ie $i \geq n+1$ in \eqref{eq:ellipsis_events} and \eqref{eq:ellipsis_log_size}.
The impact of any inaccuracies in determining $p_i$ can be minimized by
  increasing $n$, the number of sequences converted to templates.

\subsection{Memory Tradeoff}
\label{sub:memory_footprint}
The tradeoff for \Sys' benefits are computational overheads (evaluated in $\S$\ref{sec:eval_overheads} and $\S$\ref{sec:eval_overheads_scaling}) and the memory cost of storing templates ({\boldmath$M_{\tau}$}).
Let {\boldmath$M_{fixed}$} be memory required per template, excluding syscalls, while {\boldmath$M_{syscall}$} be the memory required for each syscall in the template.
On 32 bit kernel $M_{fixed} = 116$ and $M_{syscall} = 56$ bytes, determined by \textit{sizeof}  data structures.
As an example, 3 templates from evaluation occupied 2 KB in memory (Appendix~\ref{sec:appendix_templates})

\begin{equation}%
	\label{eq:runtime_mem}%
	M_{\tau} = M_{fixed} * n + M_{syscall} * \textstyle \sum_{i=1}^{n}len(s_i)%
\end{equation}%

For reference, the parameters for the application detailed in Section~\ref{sec:ardu_setup} are provided in Table~\ref{tab:ardu_eq}.
Complete templates for the same can be found in Appendix~\ref{sec:appendix_templates}.
The 3 templates used for the case study took 2 KB of memory space.

While templates occupy memory space, as per \eqref{eq:runtime_mem}, \Sys minimizes {\tt kaudit} buffer usage, reducing the memory space that must be dedicated to it.
These savings are not considered in~\eqref{eq:runtime_mem}. Section~\ref{sec:eval_buffer_util} evaluates the reduction in {\tt kaudit} buffer occupancy when using \Sys.

\subsection{Extended Reduction Horizon}
Until now we have limited the horizon of reduction to
  individual task {\it loop} instances.
We can further optimize by
  creating a single record that describes multiple consecutive matches of a template.
This higher performance system is henceforth referred to as \SystwoBold.
When a \Systwo match fails,
  a separate record is logged for each of the
  base template matches along with complete log sequence for the current instance
  (\ie the base behavior of \Sys).
\Systwo performs best when identical sequences occur continuously, capturing all sequence repetitions in one entry.

\begin{align}
	E_{\textit{\Systwo}}^{Best} &= n + I * \textstyle \sum_{i=n+1}^{N}(p_{i} * len(s_i)) + f
\\
	E_A - E_{\textit{Ellipsis-HP}}^{Best} &= I * \textstyle \sum_{i=1}^{n}(p_i * len(s_i)) -  n
\end{align}

\begin{multline}
	\tikzmarknode{a}{\highlight{Bittersweet}{$E_A - E_{\textit{Ellipsis-HP}}^{Best}$}} = 
	\tikzmarknode{b}{\highlight{NavyBlue}{I}} *
	\tikzmarknode{d}{\highlight{Fuchsia}{$\sum_{i=1}^{n}(p_i * len(s_i))$}} - 
	\tikzmarknode{c}{\highlight{OliveGreen}{$n$}}
\end{multline}
\begin{tikzpicture}[overlay,remember picture,>=stealth,nodes={align=left,inner ysep=2pt},<-]
	\path (a.north) ++ (.1,.5em) node[anchor=south west,color=Bittersweet!85] (mitext){\textsf{\footnotesize Event count reduction}};
	\draw [color=Bittersweet!85](a.north) |- ([xshift=-0.3ex,color=Bittersweet]mitext.south east);
	\path (b.south) ++ (0,-1.3em) node[anchor=south east,color=NavyBlue!85] (mitext){\textsf{\footnotesize Iterations}};
	\draw [color=NavyBlue!85](b.south) |- ([xshift=-0.3ex,color=NavyBlue]mitext.south west);
	\path (d.north) ++ (-.6,.5em) node[anchor=south west,color=Fuchsia!85] (mitext){\textsf{\footnotesize Audit events for n sequences}};
	\draw [color=Fuchsia!85](d.north) |- ([xshift=-0.3ex,color=Fuchsia]mitext.south east);
	\path (c.south) ++ (0,-1.5em) node[anchor=north east,color=OliveGreen!85] (mitext){\textsf{\footnotesize Best case \Systwo events for n sequences}};
	\draw [color=OliveGreen!85](c.south) |- ([xshift=-0.3ex,color=OliveGreen!85]mitext.south west);
\end{tikzpicture}

\vskip 7mm

\subsection{Temporal Constraints}

\Sys and \Systwo, when used to reduce the long and most frequently occurring sequences, decrease both the volume of audit events generated and the size of log that must be stored.
RTS are sensitive to time intervals between events, thus, \Sys also considers temporal checks in the template matching process ($\S$\ref{sec:template_step2} and $\S$\ref{sec:template_runtime_match}).
\Systwo adds additional checks for inter-arrival times of different task instances.
Note that the earlier discussion on log sizes assumes that temporal constraints are always met.
An evaluation of the impact of temporal constraints on log size is provided in Section~\ref{sec:eval_timing_constraints}.

%% file: sections/eval.tex
\section{Evaluation}
\label{sec:eval}

We evaluate \Sys and \Systwo using
two real time applications.
ArduPilot~\cite{ardupilot}, a safety-critical firm-deadline autopilot application.
Motion~\cite{motion}, conversely,  is a soft real time video analysis application where the application execution paths and behavior are variable depending on configuration and inputs.
With these applications
we show that our auditing systems
\ca perform \textit{lossless auditing within the application's temporal requirements}, where Linux Audit would lose audit events or violate application's safety constraints  ($\S$\ref{sec:eval_freq}),
\cb achieve high audit log volume reduction during benign activity, in most cases ($\S$\ref{sec:eval_log_reduction} and $\S$\ref{sec:motion}) (upto $97.55 \%$), even when application can exhibit various behaviors ($\S$\ref{sec:motion}),
\cc enjoy minimal computational overhead even in an artificially created worst case scenarios ($\S$\ref{sec:eval_overheads}).
Using a set of synthetic tasks we also show that the \Sys' overhead per syscall scales independent of the size of template ($\S$\ref{sec:eval_overheads_scaling}).
Unless specified otherwise ($\S$\ref{sec:motion} and $\S$\ref{sec:eval_overheads_scaling}), all evaluations were conducted using the ArduPilot application.

\subsection{Setup}
\label{sub:setup}
All measurements were conducted on 4GB Raspberry Pi 4~\cite{rpi4}
running Linux 4.19.
The RT kernel from raspberrypi/linux~\cite{rpi_rt} was used
with  additional kconfig
({\tt CONFIG\_PREEMPT\_RT\_FULL},
{\tt CONFIG\_AUDIT},
{\tt CONFIG\_AUDITSYSCALL})
enabled.
To reduce computational variability
 due to external perturbations
 we disabled power management,
 directed all kernel background tasks/interrupts to core~0 using the \textit{isolcpu} kernel argument,
 and set CPU frequency Governor to Performance~\cite{cpu_freq}.
Audit rules for capturing syscall events  were configured to match
against our benchmark application (\ie background process activity was not audited).
We set the {\tt kaudit} buffer size to 50K as any larger values led to system panic and hangs.

\subsection{ArduPilot}
\label{sec:ardu_setup}

ArduPilot is an open source autopilot application that can fully control various classes of autonomous vehicles such as quadcopters, rovers, submarines and fixed wing planes~\cite{ardupilot}. It has been installed in over a million vehicles and has been the basis for many industrial and academic projects~\cite{bin2009design,allouch2019mavsec}.
We chose the quadcopter variant of ArduPilot, called ArduCopter, as it has the most stringent temporal requirements within the application suite.
For this application the RPi4 board was equipped with a Navio2 Autopilot hat~\cite{navio2} to provide real sensors and actuator interfaces for the application.
We instrumented the application for measuring the runtime overheads introduced by auditing.
Among the seven tasks spawned by ArduPilot, we focus primarily on a task named FastLoop for evaluating temporal overheads as it includes the stability and control tasks that need to run at a high frequency to keep the QuadCopter stable and safe.

Among the syscalls observed in the trace of ArduPilot, we found that only a small subset of syscalls were relevant to forensic analysis~\cite{gt2012}:  \texttt{execve, openat, read, write, close} and \texttt{pread64}.
Upon running the template generation script on the application binary, we obtained the most frequently occuring templates for three tasks ($n=1$, for each task), consisting of 14~\texttt{write}, 16~\texttt{pread64} calls and 1~\texttt{read} call, respectively.
These templates include expected values corresponding to the file descriptor and count arguments of the syscalls as well as temporal constraints.
Templates were loaded into the kernel when evaluating \Sys or \Systwo.
Auditing was set up to audit invocations of the syscalls made by the ArduPilot application as mentioned above.
Complete templates are provided in Appendix~\ref{sec:appendix_templates}.

\input{sections/evals/log_loss}

\input{sections/evals/buffer}
\input{sections/evals/motion_table}
\input{sections/evals/log_size}

\input{sections/evals/motion.tex}

\input{sections/evals/overheads}
\input{sections/evals/overhead_scalability.tex}

\input{sections/evals/temporal}

\subsection{Summary of Results}
\Sys provides complete audit events retention while meeting temporal requirements of the ArduPilot application, with significantly reduced storage costs.
\Systwo improves this further.
The temporal constraint allows additional temporal checks, detecting anomalous latency spikes with effectively no additional log size overhead during normal operation.

%% file: sections/evals/log_loss.tex
\subsection{Audit Completeness}
\label{sec:eval_freq}

\begin{figure}[t]
		\centering
		\includegraphics[width=\linewidth,keepaspectratio]{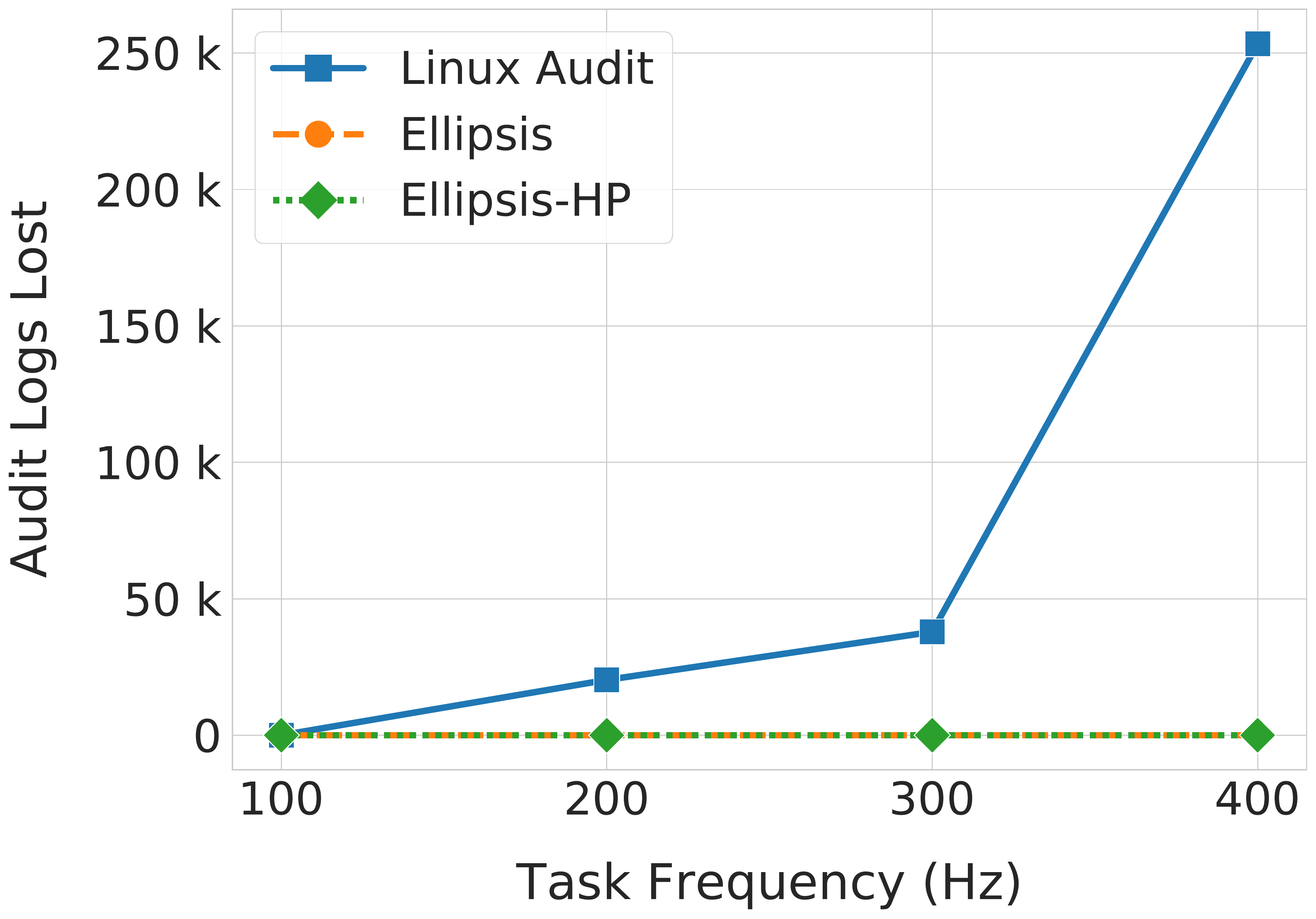}
		\caption{\label{fig:eval_freq} ($\S$\ref{sec:eval_freq})
			Number of audit events lost vs.\ frequencies of the primary loop in ArduPilot, for 100K iterations.
			 The frequency of the Fast Loop task is varied from 100~Hz to 400~Hz and the number of logs lost are plotted.
			 \Sys and \Systwo suffer no log loss
			 at any frequencies
			 and hence their lines overlap exactly with X-axis.
		}
\end{figure}

{\it Experiment.}
We ran the application for 100K iterations at task frequencies of 100 Hz, 200 Hz, 300 Hz and 400 Hz\footnote{Frequency values are chosen based on application support: \url{https://ardupilot.org/copter/docs/parameters-Copter-stable-V4.1.0.html\#sched-loop-rate-scheduling-main-loop-rate}}, measuring audit events lost.
The fast dynamics of a quadcopter benefit from the lower discretization error in the ArduPilot's PID controllers at higher frequencies~\cite{wang2020pid} leading to more stable vehicle control.

{\it Observations.}
Figure~\ref{fig:eval_freq} compares the log event loss for Linux Audit, \Sys and \Systwo across multiple task frequencies.
We observe that Linux Audit lost log events at all task frequencies above 100 Hz.
In contrast, \Sys and \Systwo did not lose audit event log at any point in the experiment.

{\it Discussion.}
Because this ArduCopter task performs critical stability and control function,
  reducing task frequency to accommodate Linux Audit may have considerable
  detrimental effects.
Further investigation ($\S$\ref{sec:eval_buffer_util}) revealed that Linux Audit dropped log events due to {\tt kaudit}
    buffer overflow, despite the buffer size being 50K.
In contrast, \Sys is able  to  provide auditing for the entire frequency range
  without suffering log event loss.

%% file: sections/evals/buffer.tex
\subsection{Audit Buffer Utilization}
\label{sec:eval_buffer_util}

\begin{figure}[t]
	\centering
	\includegraphics[width=\linewidth,keepaspectratio]{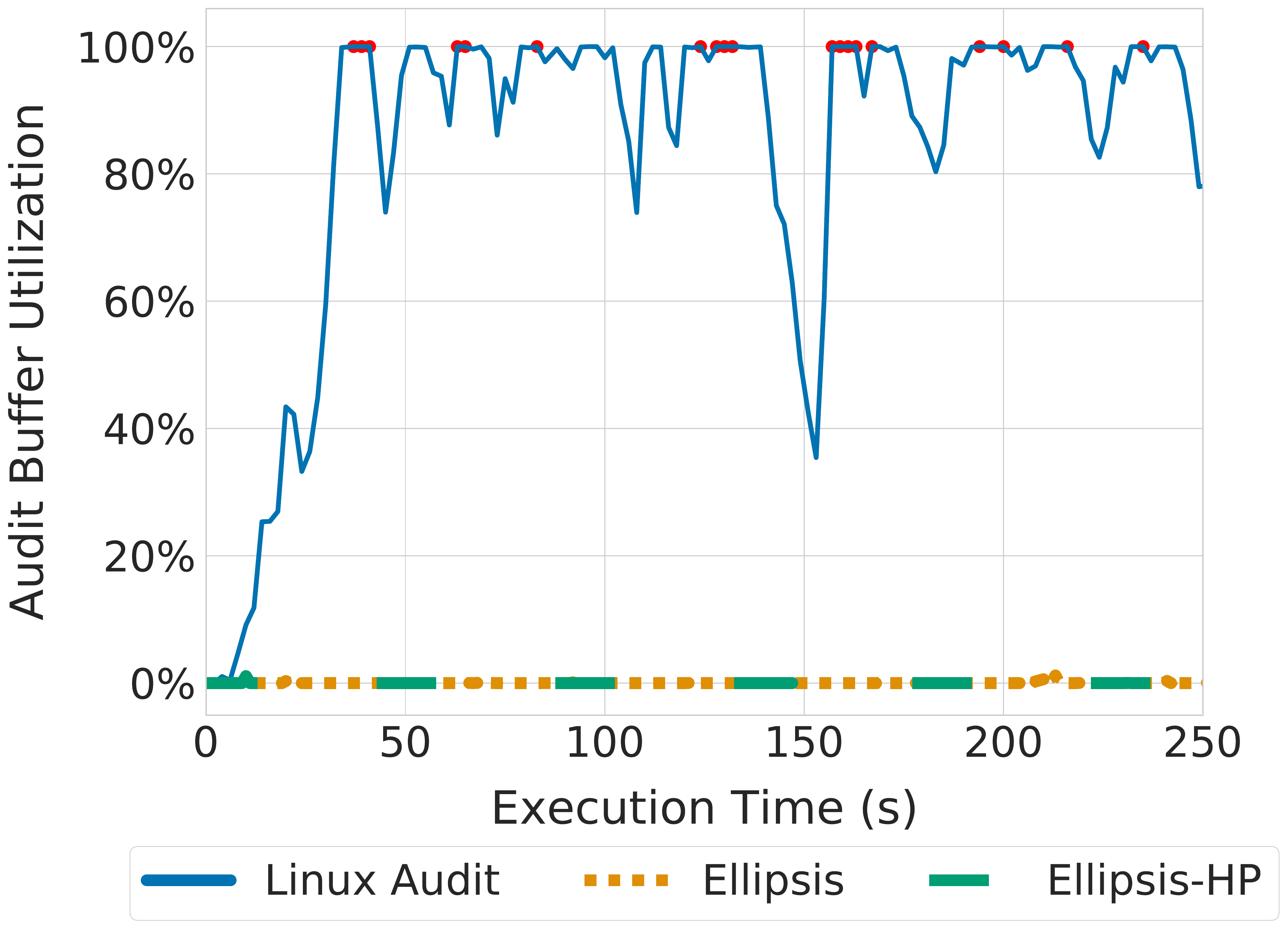}
	\caption{\label{fig:eval_backlog} ($\S$\ref{sec:eval_buffer_util})
		Audit buffer utilization.
		Additional red annotations signify times where buffer is completely filled.
		\Sys and \Systwo did not use more than 2\% buffer space.
	}
\end{figure}

{\it Experiment.}
 The size of the {\tt kaudit} buffer is determined by a ``backlog limit" configuration, that controls the number of outstanding audit messages allowed in the kernel~\cite{auditctl}.
 The default configuration is 8192 but as noted before ($\S$\ref{sub:setup}) we set it to 50K.
The kaudit buffer state was sampled periodically, once every 2 seconds, by querying the audit command-line utility \textit{auditctl} during the execution of the application for 100K iterations. Figure~\ref{fig:eval_backlog} shows the comparison of the percentage  utilization of the audit buffer by Linux Audit, \Sys and \Systwo over time.

{\it Observations.}
From Figure~\ref{fig:eval_backlog}, we see that for Linux Audit, the utilization of the kaudit buffer rises quickly and remains close to 100\% for the majority of the runtime, resulting in loss of audit messages, as measured earlier ($\S$\ref{sec:eval_freq}).
In contrast, \Sys and \Systwo ensure that the buffer utilization remains negligible throughout the execution. As noted before ($\S$\ref{sub:setup}), buffer size is already set to the largest value the platform can support without panics or hangs.


{\it Discussion.}
When the kaudit buffer is full, new audit messages are lost; hence, to ensure that suspicious events are recorded, it is essential that \textit{the buffer is never full}.
\Sys is able to keep the buffer from overflowing by reducing the number of audit logs being generated and thus reducing the number of outstanding audit logs buffered in the system.
The variations that we see in the plots can be attributed to the scheduling of the non real-time \textit{kauditd} thread that is responsible for sending the outstanding audit messages to user-space for retention on disk. We observe that the backlog builds with time when \textit{kauditd} isn't scheduled and drops sharply when \textit{kauditd} eventually gets CPU time.

However there are two limitations to using \textit{auditctl} to estimate memory usage.
First, kaudit buffer size does not consider the additional memory used by \Sys and \Systwo to maintain templates in memory and perform runtime matching.
Manual calculations yielded a memory overhead of less than 100 KB, or 1 \% of the buffer size.
Second, the relatively slow sampling rate of 0.5 Hz can miss transient changes in buffer utilization. \textit{auditctl} reports buffer occupancy at the moment it is invoked. However, running \textit{auditctl} at a higher frequency leads to changes in application profile.
So we ran further experiments to determine the minimum kaudit buffer size with which \Sys and \Systwo can still achieve complete auditing. These further experiments are free from any sampling limitation. We find that a buffer of 2.5K for \Sys and 1.5K for \Systwo was enough to support lossless auditing under normal operation.
This reduced memory requirement is valuable for RTS that run on resource constrained platforms.
The reduced time that the buffer holds audit logs, reduces the attack window for recently identified race condition attacks on the audit buffer~\cite{paccagnellalogging}.
While the buffer utilization under normal operation is vastly reduced, the buffer limit should still be kept larger than the observed minimum utilization to capture anomalous behavior without loss.



%% file: sections/evals/motion_table.tex
\definecolor{darkcandyapplered}{rgb}{0.64, 0.0, 0.0}
\newcommand{\disabled}{\color{darkcandyapplered} Disabled}
\newcommand{\Still}{\color{darkcandyapplered} Still}
\newcommand{\Limited}{\color{darkcandyapplered} Limited}

\newcommand{\enabled}{\color{ForestGreen} Enabled}
\newcommand{\Motion}{\color{ForestGreen} Motion}
\newcommand{\Verbose}{\color{ForestGreen} Verbose}

\begin{table*}[t]
    \centering
    \caption{\label{tab:motion} ($\S$\ref{sec:motion})
        Motion Application Log Reduction for different configurations
    }
        \begin{tabular}{c|c|c|c|c|c|c|c}
            \toprule
            Index &	Input   & Detection & Application Logging & Syscall Rate & Linux Audit Log Size & \Sys Log Size  & \textbf{Reduction} \\
            \midrule
            1     &	\Motion & \disabled & \Verbose            & 48.2 / s     & 7.6 MB               & 0.19 MB        & \textbf{97.55} \%  \\
            2     &	\Still  & \disabled & \Verbose            & 48.0 / s     & 7.6 MB               & 0.24 MB        & \textbf{96.80} \%  \\
            3     &	\Motion & \enabled  & \Verbose            & 44.2 / s     & 6.9 MB               & 0.29 MB        & \textbf{95.83} \%  \\
            4     &	\Motion & \disabled & \Limited            & 33.2 / s     & 5.4 MB               & 0.26 MB        & \textbf{95.21} \%  \\
            5     &	\Still  & \disabled & \Limited            & 27.6 / s     & 4.5 MB               & 0.25 MB        & \textbf{94.38} \%  \\
            6     &	\Motion & \enabled  & \Limited            & 29.7 / s     & 4.8 MB               & 0.41 MB        & \textbf{91.55} \%  \\
            7     &	\Still  & \enabled  & \Verbose            & 20.9 / s     & 3.2 MB               & 0.30 MB        & \textbf{89.83} \%  \\
            8     &	\Still  & \enabled  & \Limited            & ~8.4 / s     & 1.3 MB               & 0.24 MB        & \textbf{81.44} \%  \\
            \bottomrule
        \end{tabular}
\end{table*}

%% file: sections/evals/log_size.tex
%

\subsection{Audit Log Size Reduction}
\label{sec:eval_log_reduction}

\begin{figure}[t]
		\centering
		\includegraphics[width=\linewidth,keepaspectratio]{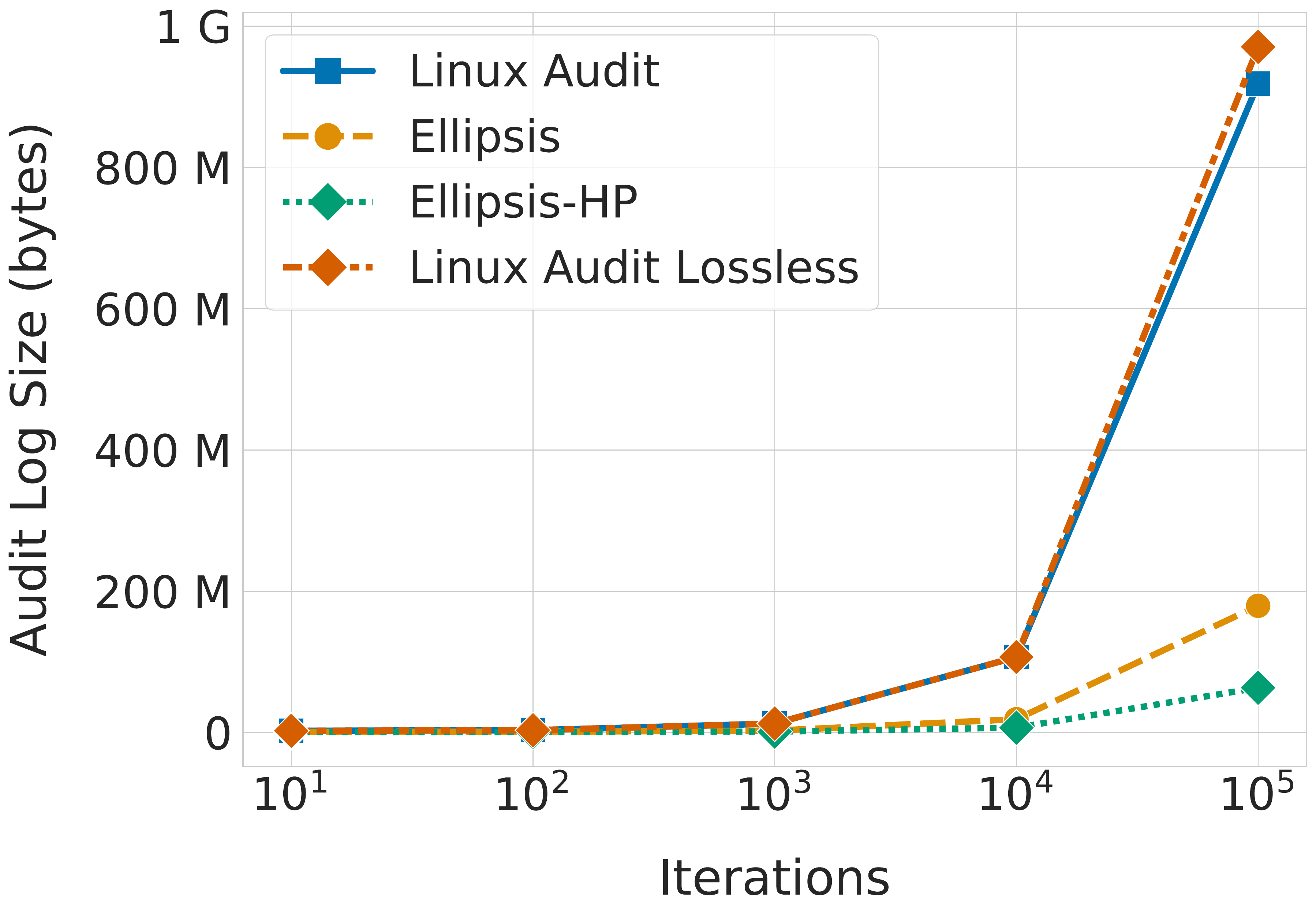}
		\caption{\label{fig:eval_reduction} ($\S$\ref{sec:eval_log_reduction})
			Total size on disk of the audit log (Y-axis), captured for different number of iterations (X-axis).
			 The size of log is measured as file size on disk in bytes.
			 $10^5$ iterations take around 250 seconds to complete.
			 For Linux Audit we measure the actual size of log on disk albeit with potential log loss.
			 Linux Audit Lossless provides an estimate of the actual size of the log on disk if the auditing was lossless.
		}
\end{figure}

{\it Experiment.}
We ran the ArduCopter application over multiple iterations in 10 to 100K range to simulate application behavior over varying runtimes.
For each iteration count, we measure the size on disk of the recorded log.

{\it Observations.}
Figure~\ref{fig:eval_reduction} compares the storage costs in terms of
  file size on disk in bytes.
The storage costs for all systems over shorter runs was found to be comparable, as the cost of auditing the initialization phase of the application ($B_A * f$) tends to dominate over the periodic loops.
Over a 250 second runtime ($10^5$ iterations) the growth of log size in \Sys was drastically lower compared to vanilla Linux Audit, with storage costs reducing by 740 MB, or \textbf{80\%}, when using \Sys. \Systwo provides a more aggressive log size reduction option by lowering storage costs by 860MB, or \textbf{93\%}, compared to Linux Audit. \textit{Linux Audit Lossless} estimates the log size had Linux Audit not lost any log events using the number of logs lost and the average size of each log entry.

{\it Discussion.}
The observations line up with our initial hypothesis that the bulk of the audit logs generated during a loop iteration would exactly match the templates. Thus, in \Sys by reducing all the log messages that correspond to a template down to a single message, we see a vast reduction in storage costs while ensuring the retention of all the audit data.
\Systwo takes this idea further by eliminating audit log generation over extended periods of time if the application exhibits expected behaviors only.
For RTS that are expected to run for months or even years without failing, these savings are crucial for continuous and complete security audit of the system.

%% file: sections/evals/motion.tex
\subsection{Motion: Audit Log Size Reduction}
\label{sec:motion}

Motion~\cite{motion} is a soft real time video analysis application where the application execution paths and behavior vary depending on configuration and inputs.

{\it Experiment.}
In this experiment, using the same setup as prior experiments, we run Motion with varying configurations and report the log reduction percentage.
Motion application detects movement in camera inputs and saves images when movement is detected within frames.
We point a camera to a screen showing a still image (Input = {\Still}) or a video\footnote{\url{https://www.youtube.com/watch?v=cElhIDdGz7M}} with random motion (Input = {\Motion}).
Motion's behavior is determined via its configuration file.
\textit{emulate\_motion} when set to on, it causes an image to be saved at a fixed rate of 2 Hz.
When \textit{emulate\_motion} is set to off, an image is stored only when motion is detected in the input stream.
Thus Motion Detection is {\enabled} when \textit{emulate\_motion=off} and {\disabled} when \textit{emulate\_motion=on}.
Application logging verbosity is set to minimum level for {\Limited} and maximum level for {\Verbose}.
For each configuration, {\color{ForestGreen}green} colored option increases execution variability compared to the other.
For each combination templates are learned over a 120 second execution.
The application is then audited with Linux Audit and \Sys for 300 seconds each.

{\it Observations.}
We provide the rate of audited syscall events, size on disk for audit logs of both Linux Audit and \Sys with the log size reduction percentage in Table~\ref{tab:motion}.
In most cases a log reduction of $> 90\%$ is achieved.
The lowest reduction occurs when the application only processes the camera feed, never saving any images.
The resultant log, with low number of syscalls and lowest size, contains disproportionately high events from the setup phase of the application, leading to a lower reduction by \Sys, which is still quite high at $81.44 \%$ .
We also experimented with doubled rate of storing images (4 Hz) but no differences were observed, as is expected.
Log loss was not observed in any scenario.

{\it Discussion.}
\Sys achieves high audit event and log reduction even when the application can have variable execution paths, $N = 26$ for this evaluation.
The only requirement is that the execution paths also be encountered in the template generation step.
The properties of repeating execution paths, shared by ArduPilot and Motion are present across a vast majority of RT applications.

%% file: sections/evals/overheads.tex
\subsection{Runtime Overheads}
\label{sec:eval_overheads}


{\it Experiment.}
This evaluation measures the execution time in microseconds ($\mu s$), for the Fast Loop task of ArduPilot, for 1000 iterations, under various auditing setups.
The small number of iterations kept the generated log volume within {\tt kaudit buffer} capacity, avoiding overflows and audit events loss in any scenario.
This avoids polluting the overhead data with instances of event loss.
The time measurement is based on the monotonic timer counter.
This process was repeated 100 times to capture the distribution of these measurements over longer application runs.
\Sys ~and \Systwo refers to the normal execution of the application with their respective reduction techniques.
To evaluate the absolute worst case for \Sys, the \Sys NR (No Reduction) scenario
  modifies the ArduCopter template so that it always fails at the last syscall.
\Sys NR is forced to fail after the longest possible partial match before eventually failing to reduce the events.
\Sys NR is also the worst case for \Systwo.

\begin{figure}[t]
	\centering
	\includegraphics[width=\linewidth,keepaspectratio]{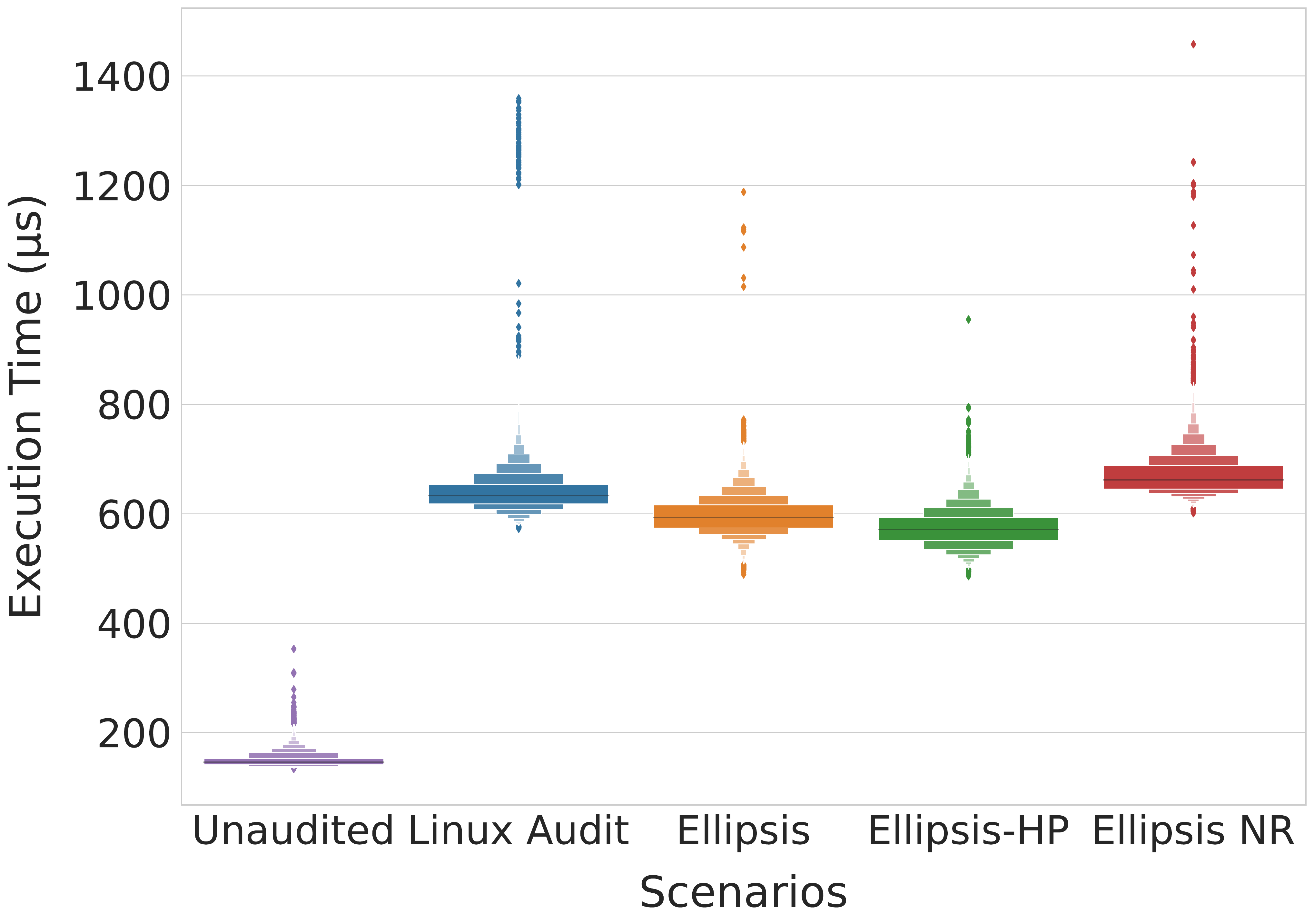}
	\caption{\label{fig:eval_overhead} ($\S$\ref{sec:eval_overheads})
		Comparison of runtime overheads of ArduPilot main loop. Task period and deadline is 2500 $\mu s$.
		Ellipsis NR (No Reduction) refers to a forced scenario where each template match fails, leading to no log reduction.
	}
\end{figure}

{\it Observations.}
Figure~\ref{fig:eval_overhead} shows the distribution of 100 execution time samples for each scenario.
  \Sys, \Systwo and \Sys NR have nearly the same overhead as Linux Audit.
  On average, \Sys's overhead is \textit{0.93}x and \Systwo's overhead is \textit{0.90}x of Linux Audit.
  The observed maximum overheads show a greater improvement.
  \Sys's observed maximum overhead is \textit{0.87}x and \Systwo's \textit{0.70}x of Linux Audit.
  \Sys NR shows a \textit{1.05}x increase in average overhead and \textit{1.07}x increase in maximum observed overhead compared to Linux Audit.

{\it Discussion.}
\Sys adds additional code to syscall auditing hooks,
  which incurs small computational overheads.
When template matches fail (\Sys NR), this additional overhead is visible,
although the overhead is not significantly worse the baseline Linux Audit.
However, in the common case where audit events are reduced by \Sys,
  this cost is masked by reducing the total amount of log collection and transmission
  work performed by Linux Audit.
This effect is further amplified in \Systwo owing to its greater reduction potential ($\S$\ref{sec:eval_log_reduction}).
Thus, \Sys's runtime overhead depends on the proportion of audit information
  reduced in the target application.
Thus, while reducing the runtime overhead of auditing is not \Sys' primary goal,
  it nonetheless enjoys a modest performance improvement by reducing the total
  work performed by the underlying audit framework.

%% file: sections/evals/overhead_scalability.tex
\subsection{Synthetic Tasks: Overhead Scaling}
\label{sec:eval_overheads_scaling}

\begin{figure}[t]
		\centering
		\includegraphics[width=.95\linewidth,keepaspectratio]{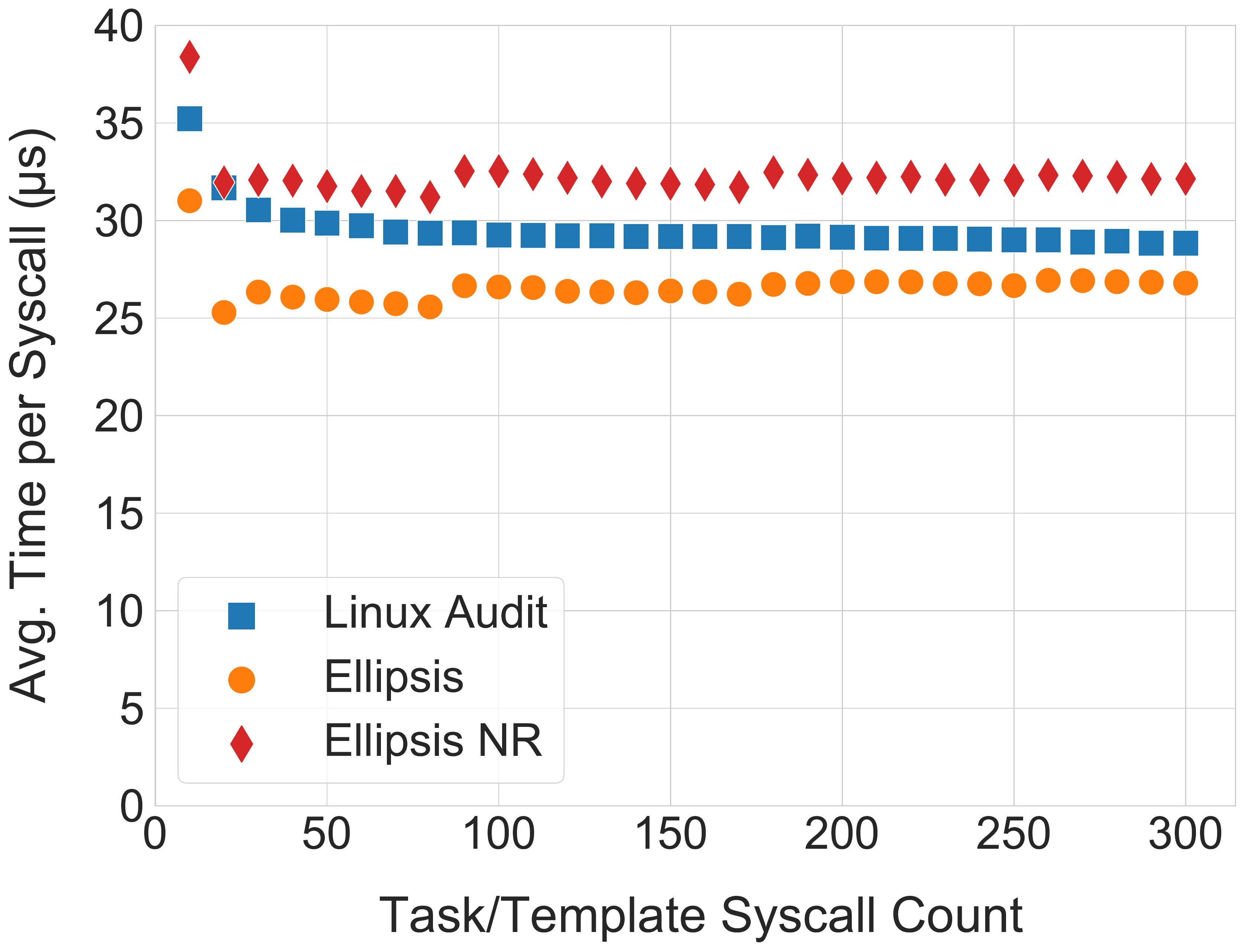}
		\caption{\label{fig:eval_scaling} ($\S$\ref{sec:eval_overheads_scaling})
			Avg. execution latency of \texttt{getpid} syscall (Y-axis) with varying task/template lengths (X-axis)
		}
\end{figure}

{\it Experiment.}
Because \Sys adds template matching logic in the critical execution path
  of syscalls, a potential concern is the overhead growth for tasks with long
  syscall sequences.
In this experiment we measure execution time for tasks that execute varying counts of \texttt{getpid} syscalls (10, 20, 30 ... 300).
\texttt{getpid} is a low latency non-blocking syscall,
  which allows us to stress-test the auditing framework.
As the max template length (\ie syscall count) observed in
  real application loops was 29,
  we analyze workloads of roughly 10 times that amount, \ie 300.
The execution time for each task is measured 100 times.
Temporal constraints are not used.
Since the tasks have a single execution path \ie a fixed count of \texttt{getpid} syscalls, \Sys' audit events reduction always succeeds.
For \Sys NR (No Reduction) we force template matches to fail at the last entry (same as $\S$\ref{sec:eval_overheads}).
This represents the worst case scenario.

{\it Observations.}
Figure~\ref{fig:eval_scaling} shows the average syscall response time
  as the number of syscalls in the task loop increases.
The primary observation of interest is that the \textit{time to execute a syscall is roughly constant}, independent of the number of syscalls in the task and template.
The higher value at the start is due to the non syscall part of the task that quickly becomes insignificant for tasks with higher number of syscalls.
We only show average latency as the variance is negligible ($< 1.3 \mu s$)%
\footnote{\label{foot:scaling_variability}
    Where \textit{isolcpu} is not used, \Sys has especially low execution time variability, detailed evaluation available as Appendix~\ref{sec:scaling_variability}}.

{\it Discussion.}
\Sys scales well as the overhead per syscall remains independent of template size, even in the worst case scenario of \Sys NR.
When log reduction succeeds the overhead is reduced.
When the log reduction fails the overhead is not significantly worse than Linux Audit.


%% file: sections/evals/temporal.tex
\subsection{Temporal Constraint Policy}
\label{sec:eval_timing_constraints}
\label{sec:temporal_contraints}


%

\begin{figure}[htp]
  \centering
  \includegraphics[width=\linewidth,keepaspectratio]{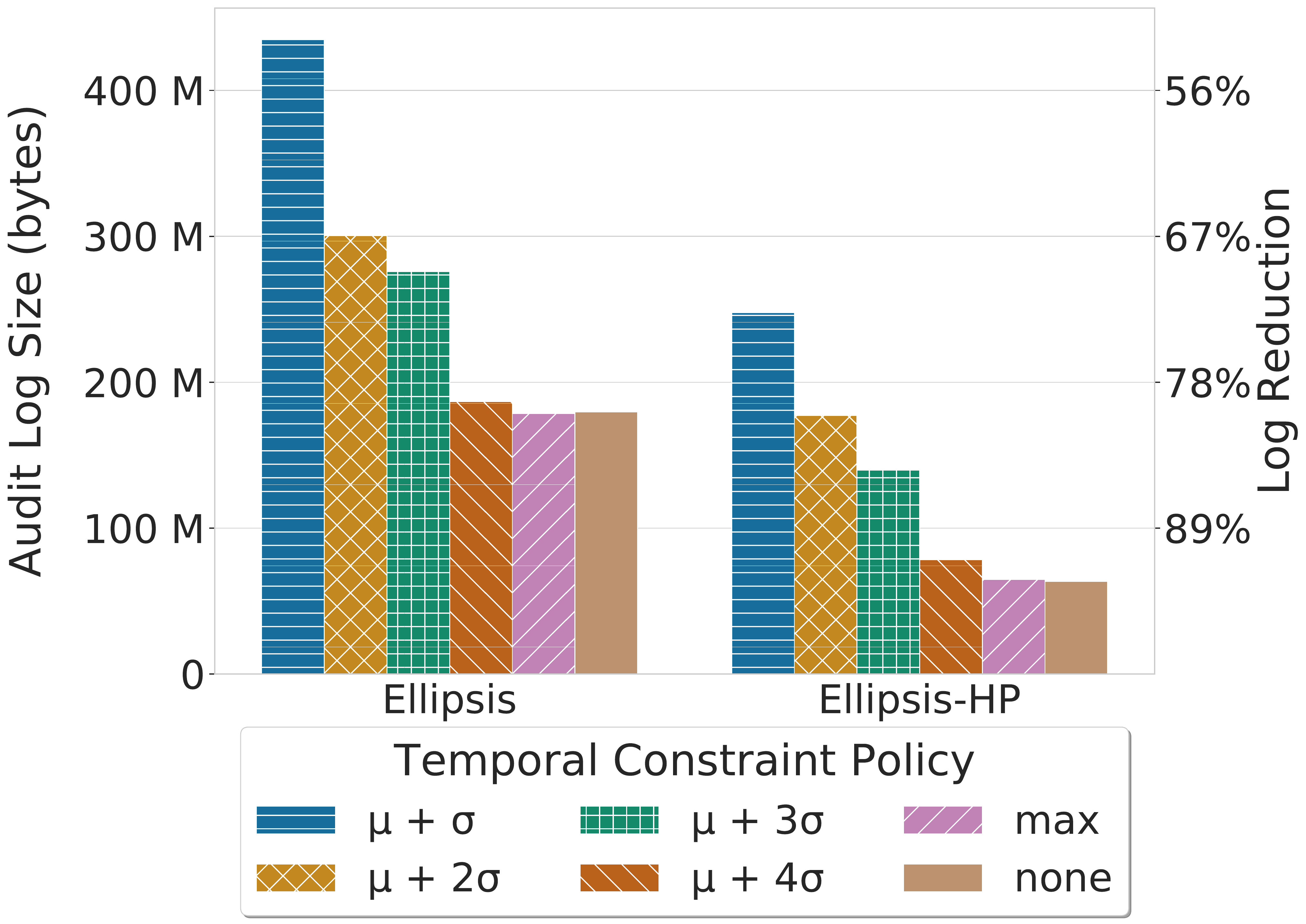}
  \caption{\label{fig:timing_constraints} ($\S$\ref{sec:eval_timing_constraints})
    Log size (Y-axis left) under varying temporal constraint policies.
    The right Y axis shows the \% reduction in log size compared to Linux Audit. 
    $\mu + 4\sigma$ covers 99.5\% of the total 100k iterations.
  }
\end{figure}

{\it Experiment.}
We explore here the impact of different policies for temporal constraints. 
Temporal constraints are applied, intra-task, for \Sys and additionally inter-task for \Systwo.
While the constraint values on expected runtimes and expected inter-arrival times of task instances are learned and applied separately for each task, a common policy can be enforced.
For example the policy \textit{max} implies that all timing constraints are set to the maximum value that was observed for them during the learning phase.
Other policies explored in this experiment are based on the average ($\mu$) and standard deviation ($\sigma$) of the time intervals observed during the learning phase. The \textit{none} policy disables all temporal constraints and represents the best case in this experiment.

{\it Observations.}
Figure~\ref{fig:timing_constraints} shows the impact of different temporal constraint policies on log size. With more stringent timing constraints, fewer task instances are observed to adhere to constraints leading to an increase in log size. \textit{max} and \textit{none} policy yield the same log size, which is expected given that temporal determinism is a design feature of RT applications.

{\it Discussion.}
The timing constraints are decided based on the observed values from the learning phase. Learning phase behavior is considered \textit{correct} as this phase is a controlled execution. Hence we believe that \Sys should be used at runtime to record unexpected behaviors \ie not seen during learning phase, while eliminating audit logs for expected behaviors. The policies \textit{max} and $\mu + 4\sigma$ most closely correspond with this recommendation.
Further is it notable that since \textit{max} and \textit{none} policy yielded almost the same log size, the \textit{max} constraint provided checks against temporal violations at a negligible cost.

%% file: sections/security_analysis_new.tex
\section{Security Analysis}
\label{sec:security_analysis}

The security goal of \Sys,
  indeed auditing in general,
  is to record all forensically-relevant information,
  thereby aiding in the investigation of suspicious activities.
The previous section established \Sys' ability to dramatically reduce audit event generation for benign activities, freeing up auditing capacity.
We now discuss the security implications of \Sys.

\subsection{Stealthy Evasion}
If a malicious process adheres to the expected behavior of benign tasks,
  the associated logs will be reduced.
The question, then, is whether a malicious process can perform meaningful actions
  while adhering to the benign templates.
If \Sys exclusively matched against syscall IDs only, such a feat may be possible;
  however, \Sys also validates syscalls' arguments and temporal constraints,
  effectively validating both the {\it control flow} and {\it data flow}
  before templatization.
Thus making it exceedingly difficult for a process
  to match a template while affecting the RTS in any meaningful way.
For example,
  an attacker might try to substitute a read from a regular file with a read from a sensitive file;
  however, doing so would require changing the file handle argument, failing the template match.
Thus, at a minimum \Sys provides comparable security to commodity audit frameworks,
  and may actually provide improved security by avoiding the common problem
  of log event loss.
A positive side effect of \Sys is built in partitioning of execution flows, benefiting provenance techniques that utilize such partitions~\cite{lzx2013-beep,mzx2016,mzw+2017}.

\subsection{Information Loss}
Another concern is whether \Sys templates
 remove forensically-relevant information.
The following is an example \texttt{write} as would be recorded by Linux Audit.
{
	\begin{lstlisting}
type=SYSCALL msg=audit(1601405431.612391366:5893333): arch=40000028 syscall=4 per=800000 success=yes exit=7 a0=4 a1=126ab0 a2=1 a3=3 items=0 ppid=1513 pid=1526 tid=1526 auid=1000 uid=0 gid=0 euid=0 suid=0 fsuid=0 egid=0 sgid=0 fsgid=0 tty=pts0 ses=1 comm="arducopter" exe="/home/pi/ardupilot/build/navio2/bin/arducopter" key=(null)
	\end{lstlisting}
}
The record above, if reduced with \Sys and reconstructed
using the \Sys log and templates, yields:
{
	\begin{lstlisting}
type=SYSCALL msg=audit([1601405431.612391356, 1601405431.612391367]:$\varnothing$): arch=40000028 syscall=4 per=800000 success=yes exit=7 a0=4 a1=$\varnothing$ a2=1 a3=$\varnothing$ items=0 ppid=1513 pid=1526 tid=1526 auid=1000 uid=0 gid=0 euid=0 suid=0 fsuid=0 egid=0 sgid=0 fsgid=0 tty=pts0 ses=1 comm="arducopter" exe="/home/pi/ardupilot/build/navio2/bin/arducopter" key=(null)
	\end{lstlisting}
}
$\varnothing$ denotes values that could not be reconstructed and [min, max] denote where a range is known but not the exact value.
Nearly all of the information in an audit record can be completely reconstructed, including
\ca all audit events executed by a task, in order of execution,
\cb forensically relevant arguments.
On the other hand, information not reconstructed is
\ca accurate timestamps,
\cb a monotonically increasing audit ID,
\cc forensically irrelevant syscall arguments.
The effect of this lost information is that fine grained inter-task event ordering and interleaving cannot be reconstructed.
This loss of information is minimal and at worst increases the size of attack graph of a malicious event.
A more detailed reconstruction example is available as Appendix~\ref{sec:appendix_reconstruct}.
We now demonstrate \Sys's ability to retain forensically relevant information.
%

\begin{figure}[t]
    \centering
    \begin{subfigure}[b]{\linewidth}
        \centering\includegraphics[width=.8\linewidth,keepaspectratio]{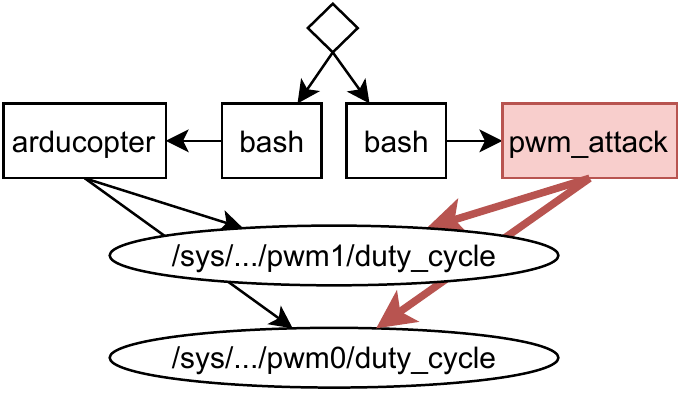}
        \caption{\label{fig:sec_attack_2} ($\S$\ref{sub:throttle_override})
             Throttle Override Attack
}
    \end{subfigure}

\vskip 7mm

    \begin{subfigure}[b]{\linewidth}
        \centering \includegraphics[width=.8\linewidth,keepaspectratio]{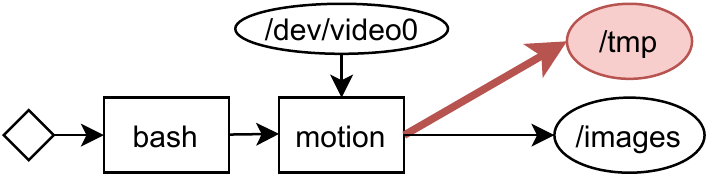}
       \caption{ \label{fig:sec_attack_1} ($\S$\ref{sub:data_exfil})
           Data exfiltration attack
       }
    \end{subfigure}
    \caption{Attack graphs created using \Sys audit logs.}
\end{figure}

\subsection{Demonstration: Throttle Override Attack}
\label{sub:throttle_override}
Autopilot applications are responsible for the safe operation of autonomous vehicles.
ArduPilot periodically updates actuation signals that control rotary speed of motors that power rotors.
The periodic updates are responsible for maintaining vehicle stability and safety. 

{\it Attack Scenario.}
Let's consider a stealthy attacker who wants to destabilize or take control over the unmanned drones.
To achieve this, the attacker first gains control of a task on the system and attempts to override the control signals.
An actuation signal's effect depends on the duration for which it controls the vehicle, therefore,
na\"ively overriding an actuation signal is not a very effective attack as the control task may soon update it to the correct value, reducing the attack's effect.
The attacker instead leverages side channel attacks such as Scheduleak~\cite{2016ScheduleBasedSA} during the reconnaissance phase of the attack to learn when the control signals are updated.
Armed with this knowledge, the attacker overrides the actuation signals immediately after the original updates, effectively taking complete control, with little computational overhead.
%
%
We use the ArduPilot setup as in described earlier ($\S$\ref{sec:ardu_setup}).
Using tools provided with Scheduleak~\cite{2016ScheduleBasedSA}, a malicious task is able to override actuation signals generated by ArduPilot.
This setup is run  for 250 seconds and audit logs collected with \Sys.


{\it Results.}
Overriding throttle control signals involves writing to files in \texttt{sysfs}.
This attack behavior can be observed in audit logs as sequences of \texttt{openat, write} and \texttt{close} syscalls.
Combining templates with the obtained audit log yields the attack graph in Figure~\ref{fig:sec_attack_2}.
\Sys correctly identifies that ArduPilot is only exhibiting benign behaviors, reducing its audit logs. 
\Sys preserves detailed attack behaviors for the malicious syscall sequences.
\Sys did not lose audit events throughout the application runtime.
In contrast, Linux Audit loses audit events ($\S$\ref{sec:eval_freq}),
potentially losing critical forensic evidence.

{\it Discussion.}
Scheduleak~\cite{2016ScheduleBasedSA} invokes \texttt{clock\_gettime} syscall frequently to infer task activation times. 
Such syscalls are irrelevant for commonly used  forensic analysis as they don't capture critical information flows. 
Despite the lack of visibility in the reconnaissance phase of the attack, auditing can capture evidence of attacker interference that creates new information flows, as shown in Figure~\ref{fig:sec_attack_2}.
We have demonstrated that when a process deviates from the expected behaviors, \eg due to an attack, \Sys provides the same security as Linux Audit.
Additionally, \Sys all but eliminates the possibility of losing portions of the malicious activity due to {\tt kaudit} buffer overflow.
However, it is impossible to guarantee that no events will ever be lost with malicious activities creating unbounded new events.
\Sys improves upon Linux Audit by
  \ca freeing up auditing resources which can then audit malicious behaviors, and
  \cb reducing the audit records from benign activities that must be analyzed as part of forensic provenance analysis.
Stealthy attacks like this also show the role of auditing in improving vulnerability detection and forensic analysis on RTS.

\input{sections/data_exfil}

%% file: sections/data_exfil.tex
\subsection{Demonstration: Data Exfiltration Attack}
\label{sec:data_exfil}
\label{sub:data_exfil}
Motion~\cite{motion} monitors camera images and detects motion by tracking pixel changes between consecutive image frames.
It is primarily used for surveillance and stores images when movement is detected. Images are stored at a location specified by the system administrator.


{\it Attack Scenario.}
The attacker inserts malicious code into the victim application to save images to an attacker controlled location when motion is detected, as shown in Figure~\ref{fig:sec_attack_1}.
The attacker can exploit another process running on the system in order to exfiltrate these images out of the system at a later point in time,
successfully leaking sensitive information.
The attack can be realized using the following code snippet, developed by Yoon~\etal~\cite{yoon2017learning} :

\begin{lstlisting}[language=C, frame=single]
    const char* orig_target_dir = cnt->conf.target_dir;
    cnt->conf.target_dir = "/tmp";
    event(cnt, EVENT_IMAGE_DETECTED, &cnt->imgs.img_ring[cnt->imgs.img_ring_out], NULL, NULL, &cnt->imgs.img_ring[cnt->imgs.img_ring_out].timestamp_tv);
    cnt->conf.target_dir = orig_target_dir;
\end{lstlisting}

{\it Experimental Setup.}  Using the same platform setup as in rest of this paper,
we ran Motion v4.3.2 using a webcam as a video source. While not commonly considered forensically relevant, we include \texttt{ioctl, rt\_sigprocmask} and \texttt{gettimeofday} in our audit ruleset as these syscalls are used to capture frames from video devices and maintain video frame rates. By running the template generation tools we obtained  two templates that describe how Motion \ci captures an image frame and  \cii captures an image frame with movement and saves it to the file system.
%
Both Motion and the malicious application are audited by \Sys for 5 minutes. We introduce movement in the camera's field of view to trigger image stores.
Images get stored to both benign and malicious locations in the system.

{\it Results.}
\Sys correctly reduces audit logs that correspond to the capture of image frames where motion is not detected because that behavior matches the templates. 
As the attacker inserts code to copy image frames describing movement, \Sys observes additional occurrences of \texttt{openat, write} and \texttt{close} syscalls that differ from behavior described by the templates,
therefore retaining complete audit logs generated in response to observed movement.



%% file: sections/discussion.tex
\section{Discussion}
\label{sec:discussion}

\subsection{System Scope \& Limitations}
\Sys is useful for any application that has predictable repeating patterns.
When sequence counts are too numerous with no high probability sequences,
  it may be possible that  too much of system memory would be required
  to achieve  significant log reduction.
That said, a large number of possible sequences is not detrimental to \Sys
    as long as there exist some high probability sequences.
\Sys's efficacy is also not dependent on specific scheduling policies unless tasks share process and thread ids;
  if task share process/thread ids and the scheduler can reorder them,
  \Sys cannot distinguish between event chains, leading to unnecessary template match failures.
  Lastly, we note that while we have motivated our design by discussing periodic
tasks, \Sys is able to work effectively on any predictable execution profiles;
for example, \Sys would also be effective for aperiodic or time table triggered
tasks, which are significantly prevalent in industrial RTS~\cite{akesson2020empirical}.

\subsection{Auditing Hard Deadline RTS}
    \Sys, like Linux Audit and Linux itself, is unsuitable for hard-deadline RTS.
    All synchronous audit components must meet the temporal requirements for Hard RTS with bounded WCET, including syscall hooks and \Sys template matching.
    Additionally the {\tt kaudit buffer} occupancy must have a strict upper bound.
    In this paper \Sys takes a long step forward, deriving high confidence empirical bounds ($\S$\ref{sec:eval_overheads}) to enable \Sys' use in firm- or soft-deadline RTS, which are prolific~\cite{akesson2020empirical}.
    However, the strict bounds required for Hard RTS are a work in progress.

\subsection{Unfavorable Conditions}
   We consider here the impact of using \Sys to audit hypothetical RTS where our assumptions about RTS properties do not hold ($\S$\ref{sec:rts_concepts}).
    If the RTS may execute previously unknown syscall sequences, extra events would exist in the audit log. 
    The audit log recorded by \Sys would thus be larger.
    Since safety, reliability and timing predictability are important requirements for RTS~\cite{akesson2020empirical} the gaps in code coverage can only be small. Hence the unknown syscall sequences will not have a major impact on audit events and log size.
    If known syscall sequences have near uniform probability of occurrence, simply using templates for them all achieves high reduction ($n = N$).
    The tradeoff is additional memory required to store templates which is a small cost (Eq.~\eqref{eq:runtime_mem}). 
    Finally, if the above are combined, sequences with substantial probability of occurrence would remain untested during the RTS development.
    For such a system, functional correctness, reliability, safety or timing predictability cannot be established, making this RTS unusable.

\subsection{Code Coverage}
\label{sec:discuss_code_coverage}
While high code coverage is important in our motivating use case
   of RT applications where reliability is a concern~\cite{chen2001effect},
   it should be noted that perfect code coverage is not a requirement for the use of \Sys.
\Sys can be deployed on any system for any application under audit,
  where the log reduction benefit is proportional to the ratio of
  runtime spent in previously analysed execution paths that are included as templates.
Perfect code coverage does allow for accurate objective analysis of \Sys' audit event and log reduction potential,
  using~\eqref{eq:sys_save_event}~and~\eqref{eq:sys_save},
  as perfect code coverage analysis would yield complete sequences $s_i$.

\subsection{Deployment Considerations}
The mechanisms for template use are fully flexible.
Any sequence for any task can be independently reduced with \Sys.
However, to use \Sys beneficially, sequences with high probability of occurrence ($p_i$) should be chosen \ie top $n$ sequences by high $p_i$ out of total $N$.
The primary trade-off is the memory cost of storing templates, as in \eqref{eq:runtime_mem}.
For an RTS with limited memory, using~\eqref{eq:ellipsis_events}~and~\eqref{eq:runtime_mem}, $n$ value can be chosen for each task independently to minimize the \Sys events generated.
The parameter $n$ is chosen independently for each task, allowing highly optimized use of main memory available for storing templates.
A second trade-off is security. As the information lost by \Sys is minimal ($\S$\ref{sec:security_analysis}), the trade-off on security is also minimal.

%% file: sections/relatedwork.tex
\section{Related Work}
\label{sec:relatedwork}

{\it System Auditing.}
Due to its value in threat detection and investigation,
  system auditing is a subject of interest in traditional systems.
While a number of experimental audit frameworks have incorporated notions
  of data provenance \cite{btb+2015,phg+2017,pmm+2012,pse+2017} and taint tracking \cite{bhc+2018,mzx2016},
  the bulk of this work is also based on commodity audit frameworks such as Linux Audit.
Techniques have also been proposed to efficiently extract
  threat intelligence from voluminous log data \cite{cgh+2019,gxl+2018,hal+2018,hgl+2019,hmw+2017,hnd+2020,hpb+2020,kks+2016,kww+2018,lzl+2018,meg+2018,mge+2019,mzw+2017,pgs+2016,phm+2018,tzl+2018,whl+2020,sundaram2012prius};
  in this work, we make the use of such techniques applicable to RTS through the design of a system audit framework that is compatible with temporally constrained applications.
Our approach to template generation in \Sys shares similarities with the notion of {\it execution partitioning} of log activity \cite{lzx2013-beep,kww+2018,mzw+2017,hgl+2019,hnd+2020}, which decomposes long-lived applications into autonomous units of work to reduce false dependencies in forensic investigations.
Unlike past systems, however, our approach requires no application instrumentation
  to facilitate.
Further, the well-formed nature of real-time tasks ensures
  the correctness of our execution units \ie templates.

{\it Auditing RTS.}
Although auditing has been widely acknowledged as an important aspect of securing embedded devices \cite{embedded_audit_2,embedded_audit_3,embedded_audit_1},
challenges unique to auditing RTS have received limited attention.
Wang \etal present ProvThings, an auditing framework for monitoring IoT smart home deployments \cite{whb+2018},
but rather than audit low-level embedded device activity their system monitors API-layer flows on the IoT platform's cloud backend.
Tian et al. present a block-layer auditing framework for portable USB storage  that can be used to diagnose integrity violations \cite{tbb+2016}.
Their embedded device emulates a USB flash drive,
but does not consider syscall auditing of RT applications.
Wu et al. present a network-layer auditing platform that captures the temporal properties of network flows and can
thus detect temporal interference \cite{wcp+2019}.
Whereas their system uses auditing to diagnose performance problems in networks,
the presented study considers the performance problems created by auditing within RT applications.

{\it Forensic Reduction.} Significant effort has been dedicated to improving the cost-utility ratio for
  system auditing by pruning, summarizing, or otherwise
  compressing audit data that is unlikely to be of use during investigations
  \cite{bbm2015,bth+2017,bhc+2018,hal+2018,lzx2013,mlk+2015,tll+2018,xwl+2016,sundaram2012prius,chen2017distributed,hws+2018}.
However these approached address the log storage overheads and not the voluminous event generation that is prohibitive to RTS auditing ($\S$\ref{sec:eval_freq}).
KCAL~\cite{mzk+2018} and ProTracer \cite{mzx2016} systems are among the few that, like \Sys, inline their reduction methods into the kernel.
Regardless of their layer of operation, these approaches are often based on an observation that certain log semantics
  are not forensically relevant (\eg temporary file I/O \cite{lzx2013}),
  but it is unclear whether these assumptions hold for real-time cyber-physical environments,
\eg KCAL or ProTracer would reduce multiple identical reads syscalls to a single entry.
However, a large number of extra reads can cause catastrophic deadline misses.
Forensic reduction in RTS, therefore, needs to be cognizant of the characteristics of RTS or valuable information can be lost.
Our approach to template generation in \Sys shares similarities with the notion of {\it execution partitioning} of log activity
\cite{hgl+2019,hnd+2020,kww+2018,lzx2013-beep,mzw+2017},
which decomposes long-lived applications into autonomous units of work to reduce false dependencies in forensic investigations.
Unlike past systems, however, our approach requires no instrumentation to facilitate.
Further, leveraging the well-formed nature of real-time tasks ensures
  the correctness of our execution units \ie templates.
To our knowledge, this work  is the first to address the need for forensic reduction of system logs in RTS.

%% file: sections/conclusion.tex
\section{Conclusion}
\label{sec:conclusion}
\Sys is a novel audit event reduction system that exemplifies synergistic application-aware co-design of security
mechanisms for RTS.
\Sys allows RT applications to be audited while meeting the temporal requirements of the application.
The role of auditing in securing real-time applications can now be explored and enhanced further.
As showcased with Auditing in this work, other security mechanisms from general purpose systems warrant a deeper analysis for their use in RTS.

%% file: sections/timing.tex
\section{Audit Framework Analysis}
\label{sec:basetiming}
\label{sec:appendix_basetiming}

The Linux Audit system provides a way to observe and analyze system activities.
While Linux Audit can be configured to monitor high-level activities such as login attempts, its primary utility (and overhead) comes from tracking low-level syscalls, which is the focus of this work.

\subsection{Setup}
We use a 4GB Raspberry Pi 4~\cite{rpi4}
running Linux 4.19.
The kernel from raspberrypi/linux~\cite{rpi_rt}
was used
 with kconfig
({\tt CONFIG\_PREEMPT\_RT\_FULL},
  {\tt CONFIG\_AUDIT},
  {\tt CONFIG\_AUDITSYSCALL})
   enabled.
   To reduce computational variability due to external perturbations we disable power management, direct all kernel background tasks/interrupts to core~0 using the \textit{isolcpu} kernel argument and set CPU frequency Governor to Performance~\cite{cpu_freq}.
 Audit rules for capturing syscall events  were configured to only match
   against our benchmark application (\textit{i.e.}, background process activity was not audited).
We set the {\tt kauditd} buffer size to 50K as it was found to be the highest stable configuration possible for the evaluation platform. Larger values led to system panic and hangs.

\begin{figure*}[t]
    \begin{minipage}[t]{0.48\linewidth}
        \centering
        \includegraphics[width=\linewidth,keepaspectratio]{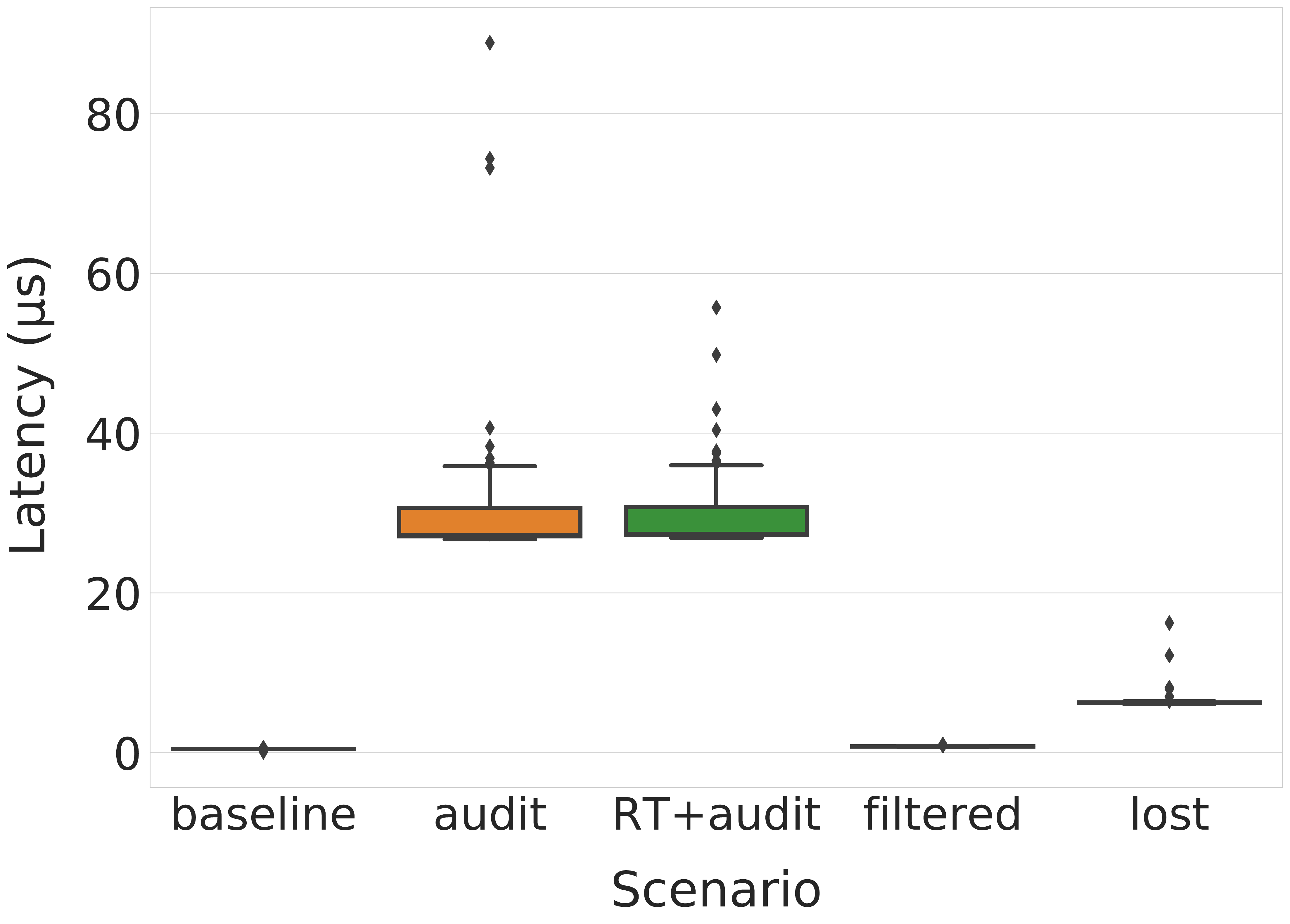}
        \caption{\label{fig:overheads}
            Latency distribution of executing  \texttt{getpid} for various auditing scenarios.
        }

    \end{minipage}
    \hfill
    \begin{minipage}[t]{0.48\linewidth}
        \centering
        \includegraphics[width=\linewidth,keepaspectratio]{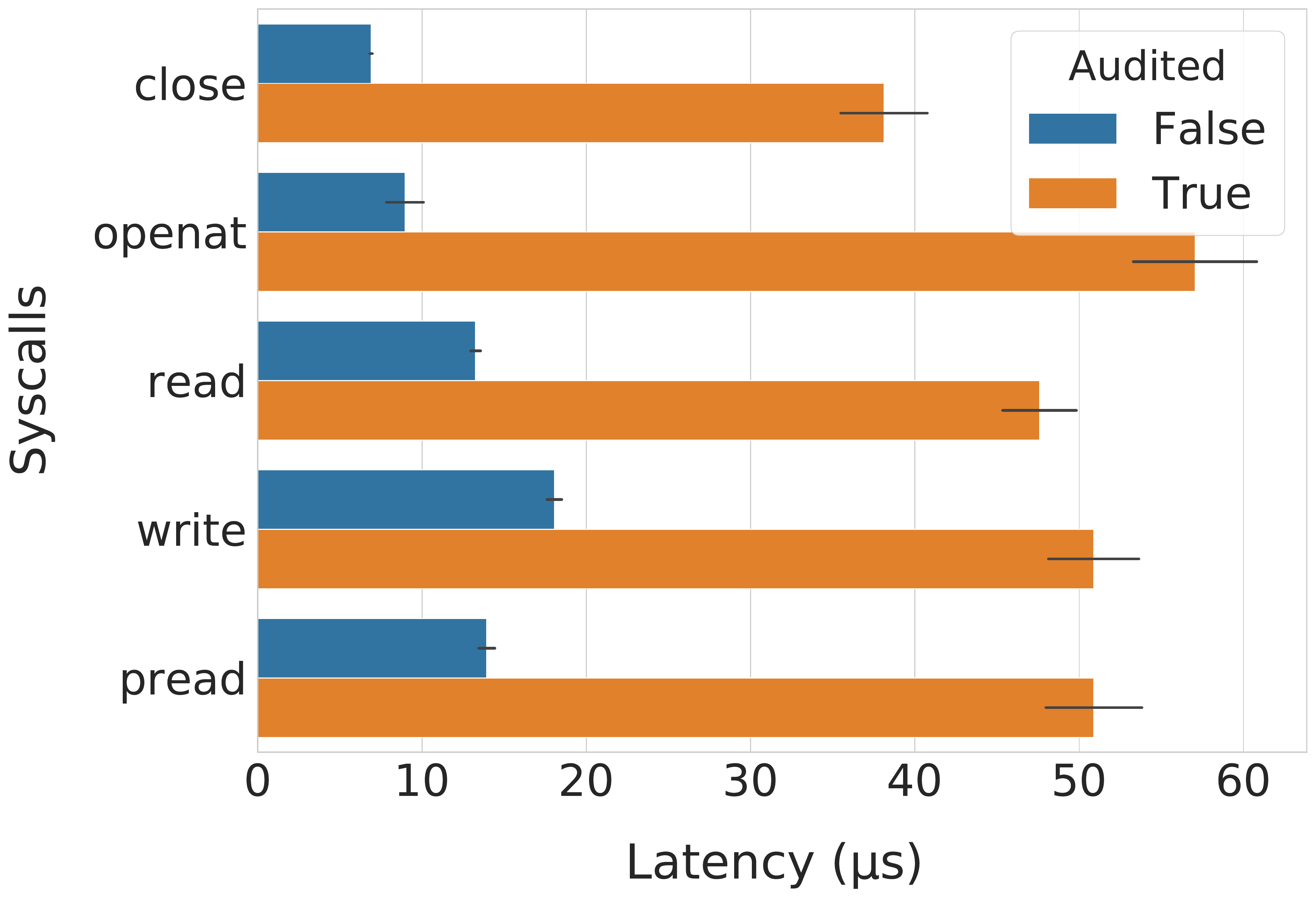}
        \caption{\label{fig:syscall_overhead}
            Execution times of syscalls. 
            The bars show mean latency over 1000 iterations, with whiskers denoting $\sigma$.
        }
    \end{minipage}
\end{figure*}

A microbenchmark, code below, is executed to observe variations in syscall execution time in the presence of external factors such as auditing, real time scheduling priorities, background stress and parallel execution.
Since the goal of this analysis is to measure the overheads imposed on syscalls
  by the audit framework, it is necessary to minimize the latency of
  the syscall itself.
For this reason, our analysis primarily uses \texttt{getpid}, a low latency
  non-blocking syscall that does not require any arguments.
Other syscalls were also evaluated to ensure the generality of observations ($\S$\ref{sec:temporal_1}, Fig.~\ref{fig:syscall_overhead}).



    \lstset{
        tabsize=2,
        showspaces=false,
        float=[htb],
        basicstyle=\footnotesize,
    }
\begin{lstlisting}[language=C, frame=single]
for (i = 0; i < 1000; i++) {
  clock_gettime(CLOCK_MONOTONIC,start_data);
  syscall(); // Replaced with specific syscalls
  clock_gettime(CLOCK_MONOTONIC,stop_data);
  clock_gettime(CLOCK_MONOTONIC,empty_start);
  clock_gettime(CLOCK_MONOTONIC,empty_stop);
  latency = timespec_subtract(start_data, stop_data) -
            timespec_subtract(empty_start, empty_stop);}
\end{lstlisting}

\subsection{Temporal Analysis}
\label{sec:temporal_1}
\textit{Experiment:}
We first measure the overhead added by Linux Audit when processing an individual syscall.
Figure~\ref{fig:overheads} shows the latency to execute the benchmark application
  issuing a \texttt{getpid} syscall.
Each column shows a box plot of \texttt{getpid} execution latency over 1000 iterations.
The \textit{baseline} scenario has auditing disabled and no other application running.
For the  \textit{audit} scenario, the baseline is repeated but the benchmark application is under audit for the {\tt getpid} syscall..
For the \textit{RT+audit} scenario, we execute the previous scenario with the
  benchmark application running at an RT priority.
In the {\it filtered} scenario, the benchmark application is still under audit but
  {\tt getpid} is no longer in the ruleset.
In all earlier scenarios, the {\tt kauditd} buffer never overflows and
  complete logs are captured;
  the {\it lost} scenario shows the latency imposed on {\tt getpid} when
  the audit framework attempts to create a log event but is unable to due to
  buffer overflow, causing the event to be lost.

\textit{Observations:}
As can be seen, the observed maximum overhead was just under 100 $\mu$s ({\it audit}).
This max reduces to 60 $\mu s$ when the application under audit is assigned an RT priority.
If a syscall is excluded from audit rules or the event log is lost due to the buffer being full, the overhead is much smaller at $\leq 1 \mu s$ and $\leq 20 \mu s$, respectively.
Figure~\ref{fig:syscall_overhead} shows the latencies for \textit{baseline}
(Audited = False)
and \textit{RT + audit}
(Audited = True)
scenarios for different syscalls.
These latencies appear to hold roughly consistent for different syscalls.

\begin{figure*}[t]
	\begin{minipage}[t]{0.48\linewidth}
		\centering
		\includegraphics[width=\linewidth,keepaspectratio]{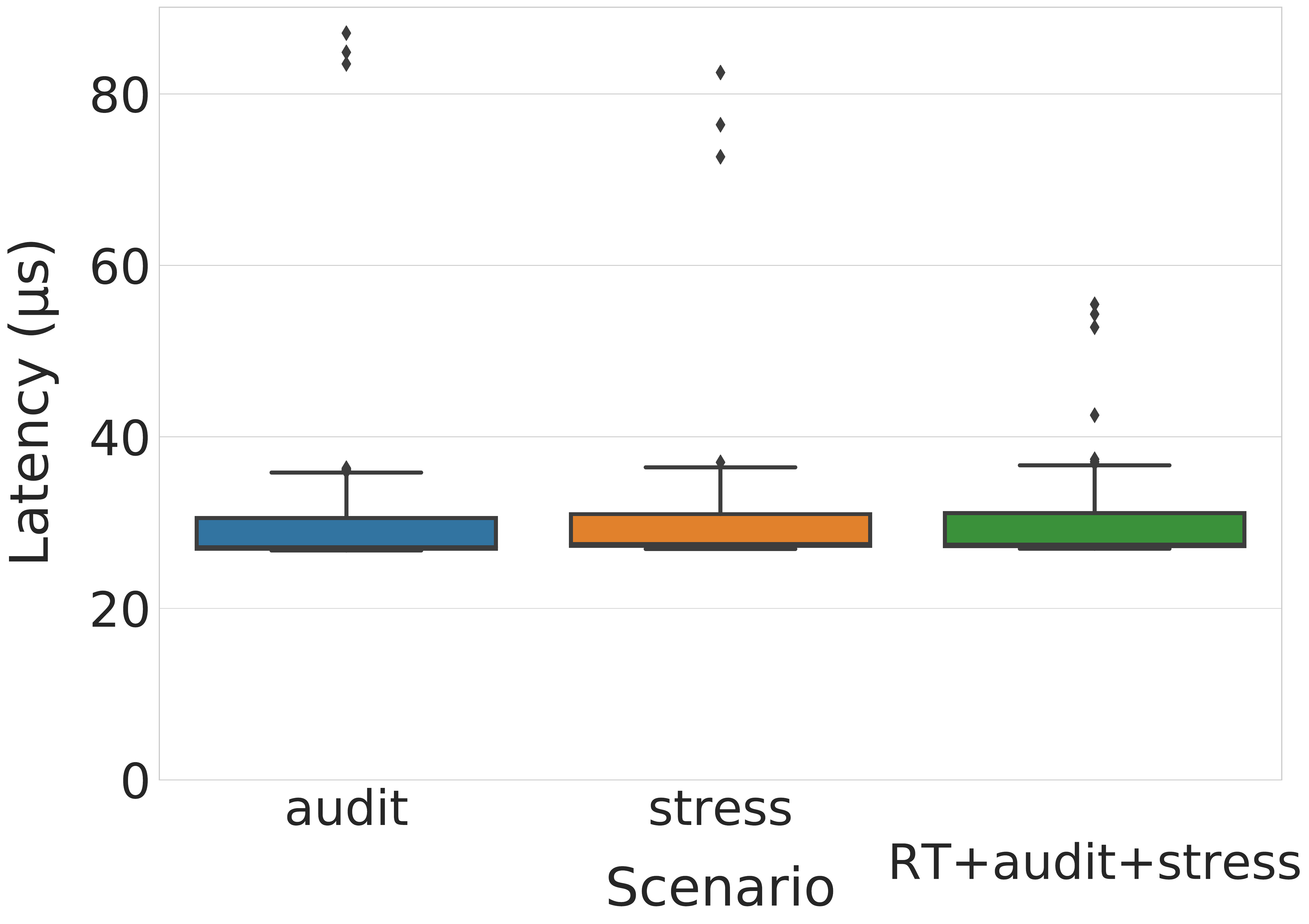}
		\caption{\label{fig:overheads_stress}
			Overhead of auditing the \texttt{getpid} benchmark application with background stress.
		}

	\end{minipage}
	\hfill
	\begin{minipage}[t]{0.48\linewidth}

    \centering
    \includegraphics[width=\linewidth,keepaspectratio]{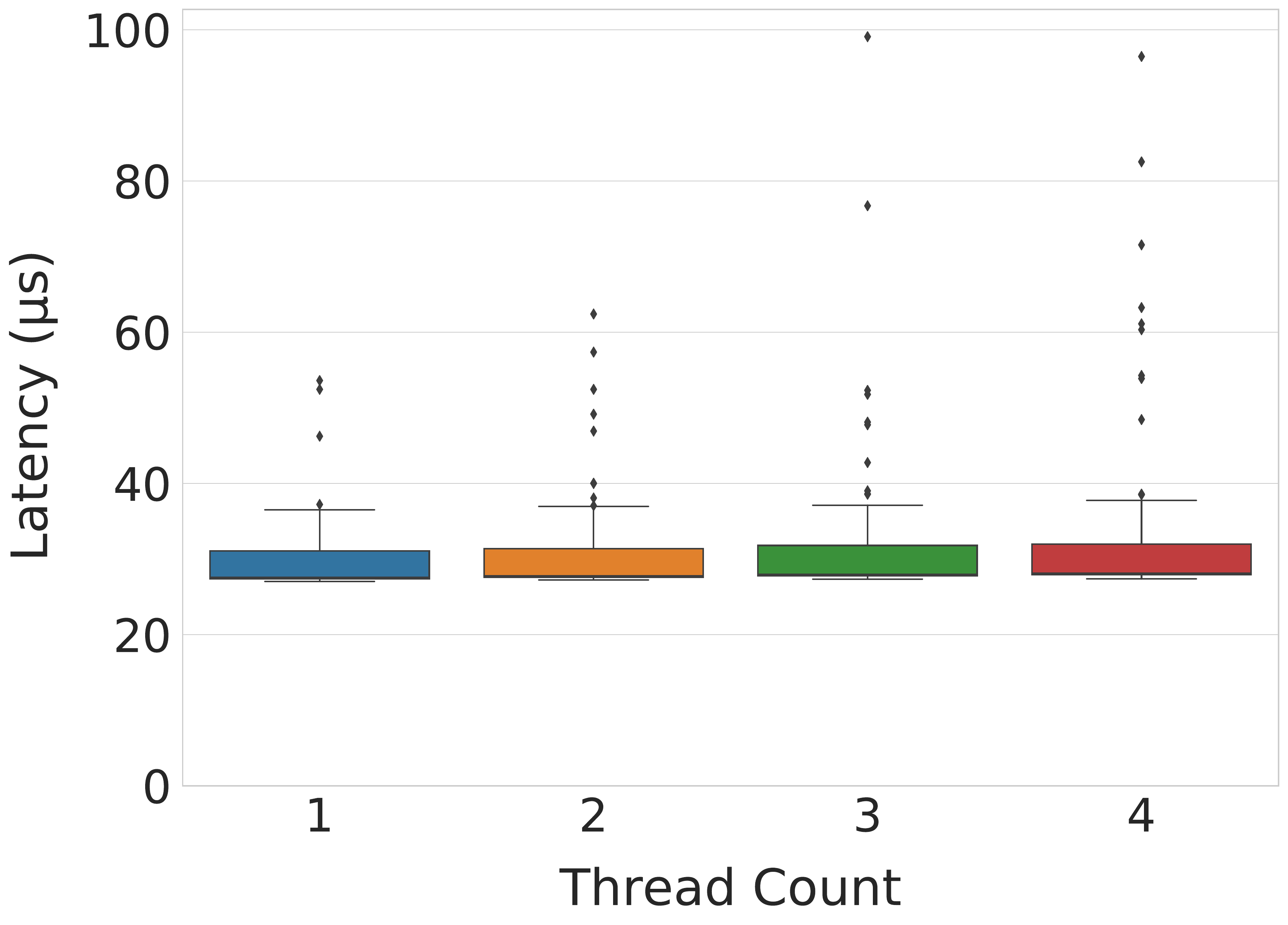}
    \caption{\label{fig:contention}
        Latency of a syscall execution with an increasing number of parallel threads executing at same priority.
    }
    	\end{minipage}
\end{figure*}

\subsection{Priority Inversion}
\label{sub:priority_inversion}
\textit{Experiment:}
Priority inversion~\cite{priority_inversion}, an instance of resource contention
 in which a higher priority task is blocked by a lower priority task,
 is a serious threat to the stability of RTS.
In the case of system auditing, a potential concern is that a low priority task's
  usage of auditing system could block a higher priority task by the virtue of shared usage of auditing resources and services
  \eg backlog buffer and audit daemons.
To investigate this,
  we repeat the {\it RT+audit} scenario from the past experiment,
  introducing a stress application running parallel in the background.
The stress application is audited and executes the same \texttt{getpid}
  workload over 4 threads, {\it but is run at a lower non-RT priority.}
Figure~\ref{fig:overheads_stress} reports the latency of syscall execution {\it for the main benchmark application only}, \ie not including the stress application's latencies.

\textit{Observations:}
 The \textit{RT+audit+stress} scenario in Figure~\ref{fig:overheads_stress}  is equivalent to the \textit{RT+audit} scenario in Figure~\ref{fig:overheads}, except with the addition of the stress workload.
 We find that when an audited application runs with high priority it does not suffer any additional contention from any non-real-time audited workloads being run in parallel, as we observe negligible latencies compared to the baseline.
 Hence, the audit framework does not
 induce priority inversion when auditing applications with different priority.
This is an expected (albeit reassuring) result because the insertion of new logs
 into the kaudit buffer does not block and audit daemons are not run at RT priority.


\subsection{Resource Contention}
\label{sub:contention}
 \textit{Experiment:}
Linux Audit introduces a resource shared among RT tasks being audited, the {\tt kaudit} buffer protected by a spinlock.
Parallel accesses can lead to contention and blocking, even between tasks with same RT priority.
To test the presence of contention,
  we repeat the earlier benchmark in which
  a single-threaded application issues {\tt getpid} calls.
While measuring the latency of one thread,
  we introduce an increasing number of additional threads running the same benchmark at same RT priority.
The threads are synchronized via a barrier to
  start executing syscalls at the same time.
 We provision the {\tt kauditd} buffer  such that
   it is large enough to prevent overflow.
The benchmark has a tight loop that runs only the \texttt{getpid} syscall in each thread,
  hence ruling out the processor cache or memory bandwidth as sources of contention.

\textit{Observations:}
The execution times for the syscalls from the thread under observation is shown in Figure~\ref{fig:contention}.
In the average case, we observe only a small
  difference in the latency of {\tt getpid} regardless of the parallel workloads.
While the observed worst case overhead is greater with 3 or 4  threads, it is still under 100 $\mu$s even when the tasks on all 4 cores are being audited.
 Delays due to contention would occur if
   multiple threads try to access the spinlock
   at the same time; but even with the fast
   {\tt getpid} call the threads minimally contend on the spinlock.
This result intuitively follows as the shared spinlock covers a small critical section containing fast pointer manipulations only, making contention uncommon even in this unfavorable scenario with repeated calls to a fast syscall.

\subsection{Remarks}
Encouragingly, we observe that Linux Audit
  does not introduce significant issues of priority inversion or contention. Hence it is a good candidate for RTS.
While contention is possible due to the spinlock on the {\tt kauditd} buffer, this cost does not impact the average latency of auditing as the number of parallel threads increases.
Further, except for limited outlier cases, \textit{the latency introduced by syscalls can be measured and bounded}.
This works well for the latency-sensitive RT systems that RT Linux is intended for.
However, the storage and management of these audit logs remains a significant challenge for resource-bound devices \eg our experiments in $\S$\ref{sub:contention} generated over 2~MB of audit logs in 50~ms.

%% file: sections/appendix_tpl.tex
\section{Templates for ArduPilot}
\label{sec:appendix_templates}

\begin{table}[htb]
\centering
\scriptsize
\caption{\label{tab:templates}
	Columns of this table describe the three template files for ArduPilot.}
 \resizebox{\columnwidth}{!}{%
\begin{tabular}{llll}
\toprule
\textbf{Thread/Task}                                                              & arducopter & ap-rcin  & ap-spi-0 \\
\midrule
\textbf{Syscall Count}                                                                    & 14         & 16       & 1        \\
\textbf{\begin{tabular}[c]{@{}l@{}}Expected runtime (ns) \end{tabular}} & 1303419    & 671567   & 0        \\
\textbf{\begin{tabular}[c]{@{}l@{}}Expected inter-arrival time (ns) \end{tabular}} & 5012313    & 20029121 & 2010477  \\
\midrule
\textbf{Syscall List} &
  { 
  \begin{tabular}[c]{@{}l@{}}4:3:-1:1:-1\\ 4:4:-1:1:-1\\ 4:5:-1:1:-1\\ 4:6:-1:1:-1\\ 4:7:-1:1:-1\\ 4:8:-1:1:-1\\ 4:9:-1:1:-1\\ 4:10:-1:1:-1\\ 4:11:-1:1:-1\\ 4:12:-1:1:-1\\ 4:13:-1:1:-1\\ 4:14:-1:1:-1\\ 4:15:-1:1:-1\\ 4:16:-1:1:-1\end{tabular}
  }
  &
  { 
  \begin{tabular}[c]{@{}l@{}}180:17:-1:11:-1\\ 180:18:-1:11:-1\\ 180:19:-1:11:-1\\ 180:20:-1:11:-1\\ 180:21:-1:11:-1\\ 180:22:-1:11:-1\\ 180:23:-1:11:-1\\ 180:24:-1:11:-1\\ 180:25:-1:11:-1\\ 180:26:-1:11:-1\\ 180:27:-1:11:-1\\ 180:28:-1:11:-1\\ 180:29:-1:11:-1\\ 180:30:-1:11:-1\\ 180:31:-1:11:-1\\ 180:32:-1:11:-1\end{tabular}
  }
  &
  { 
  3:55:-1:8:-1
  }
  \\
  \bottomrule
\end{tabular}
 }
\end{table}

ArduPilot yielded 3 templates. 
System call numbers and their corresponding arguments, a0 - a4, were extracted from the audit logs. \texttt{read}, \texttt{write}, \texttt{pread64} have syscall numbers 3,4 and 180 respectively. Argument values of -1 and temporal constraint values of 0 denote that these arguments are ignored.
\textbf{4:3:-1:1:-1} then indicates a \textit{write syscall} with\textit{ a0 as 3, a2 as 1}. \textit{a1 and a3} are \textit{not forensically relevant}.
Table~\ref{tab:templates} describes the complete templates.
%
%
An execution sequence matching a template is reduced to a single line in the audit logs at runtime as shown in the following example
{\small
\begin{lstlisting}
type=SYSCALL msg=audit(1601405431.612391356:5893330): arch=40000028  per=800000 template=arducopter rep=10 stime=1601405431589320747 etime=1601405431612287042 ppid=1208 pid=1261 tid=1261 auid=1000 uid=0 gid=0 euid=0 suid=0 fsuid=0 egid=0 sgid=0 fsgid=0 tty=pts0 ses=3 comm="arducopter" exe="/home/pi/ardupilot/build/navio2/bin/arducopter" key=(null)\end{lstlisting}
}

Some fields in \Sys log are distinct from standard Linux Audit logs
\begin{itemize}
    \item \textit{template} : The name of the template. This is the first line of a template file \eg [arducopter, ap-rcin, ap-spi-0]  in Table~\ref{tab:templates}.
    \item \textit{rep} : The number of consecutive repetitions of the template this entry represents. For \Sys the \textit{rep} value is always 1. 
    \item \textit{stime} : Timestamp of the first syscall in this reduced sequence, unit is nano seconds.
    \item \textit{etime} : Timestamp of the last syscall in this reduced sequence, unit is nano seconds.
\end{itemize}

%% file: sections/scaling_variability.tex
\section{Overhead Variability}
\label{sec:scaling_variability}

\begin{figure}[ht]
    \centering
    \includegraphics[width=\linewidth,keepaspectratio]{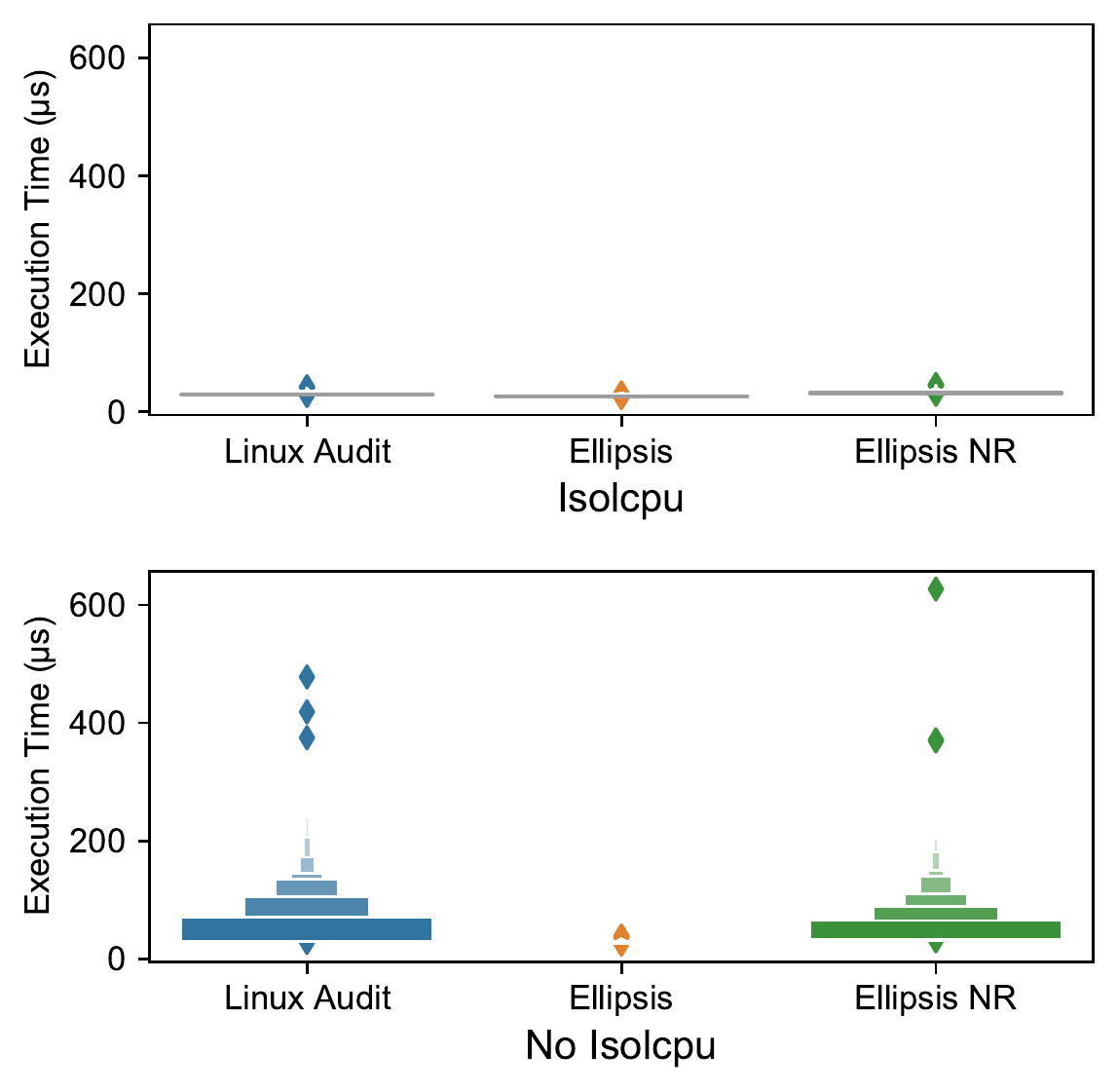}
    \caption{\label{fig:scaling_variability} 
        Syscall execution time variability.
    }
\end{figure}

In all other experiments in this work, we direct kernel background tasks and interrupts away from the cores running real time applications using the \textit{isolcpu} kernel argument, as described earlier in system setup.
The same setup was used to determine the \Sys overhead scaling.

In Figure~\ref{fig:scaling_variability} the Isolcpu case is based on the same result, arranged to show the variability introduced in the latency to execute a {\tt getpid} syscall, in the combined data of all task lengths.
The Y-axis in Figure~\ref{fig:scaling_variability} thus shows the latency to execute tasks of length $\in [10, 20, 30 ... 300]$ divided by the length of the task.
Therefore Figure~\ref{fig:scaling_variability}, Isolcpu case shows that execution latency of a syscall, average and variance, is independent of task length or template length or use of \Sys.

While isolation techniques are typically employed in multicore RTS~\cite{7176036}, there can exist RTS where due to resource constraints, kernel background processes and interrupts cannot be diverted away from the cores running real time applications.
The No Isolcpu case in Figure~\ref{fig:scaling_variability} shows the latency to execute a {\tt getpid} syscall for the same setup as before but without the use {\tt isolcpu} kernel command line argument.
\Sys significantly reduces the variability when log reduction succeeds.

Latency variability is directly related to the number of audit logs generated as only when a log is generated do the auditing hooks in syscall code paths interact with log handling daemons of Linux Audit via the kaudit buffer.
\Sys's reduction of log generation therefore also reduces variability in a syscall execution time by reducing the number of logs generated.
This benefit of \Sys can be valuable to RTS that cannot dedicate a cpu core to interrupts and background processes.

%% file: sections/appendix_reconstruct.tex
\section{Audit Log Reconstruction}
\label{sec:appendixB}
\label{sec:appendix_reconstruct}

This section shows how information can be constructed back from \Sys output.
It also notes what specific information is lost in the compression decompression process. For sake of brevity this example shows a simplified stencil of length 3 and only considers \Sys.

Let's assume that three events are recorded at runtime that would have generated the following log without \Sys:
{
\begin{lstlisting}
type=SYSCALL msg=audit(1601405431.612391356:5893330): arch=40000028 syscall=4 per=800000 success=yes exit=8 a0=3 a1=126aa4 a2=1 a3=3 items=0 ppid=1513 pid=1526 tid=1526 auid=1000 uid=0 gid=0 euid=0 suid=0 fsuid=0 egid=0 sgid=0 fsgid=0 tty=pts0 ses=1 comm="arducopter" exe="/home/pi/ardupilot/build/navio2/bin/arducopter" key=(null)
type=SYSCALL msg=audit(1601405431.612391366:5893333): arch=40000028 syscall=4 per=800000 success=yes exit=7 a0=4 a1=126ab0 a2=1 a3=3 items=0 ppid=1513 pid=1526 tid=1526 auid=1000 uid=0 gid=0 euid=0 suid=0 fsuid=0 egid=0 sgid=0 fsgid=0 tty=pts0 ses=1 comm="arducopter" exe="/home/pi/ardupilot/build/navio2/bin/arducopter" key=(null)
type=SYSCALL msg=audit(1601405431.612391367:5893334): arch=40000028 syscall=4 per=800000 success=yes exit=7 a0=5 a1=126ab8 a2=1 a3=3 items=0 ppid=1513 pid=1526 tid=1526 auid=1000 uid=0 gid=0 euid=0 suid=0 fsuid=0 egid=0 sgid=0 fsgid=0 tty=pts0 ses=1 comm="arducopter" exe="/home/pi/ardupilot/build/navio2/bin/arducopter" key=(null)\end{lstlisting}
}

Let's further assume that following template was loaded for the process.
{
\begin{lstlisting}
arducopter
3
1303419
5012313
4:3:-1:1:-1
4:4:-1:1:-1
4:5:-1:1:-1\end{lstlisting}
}

\Sys compresses the three events into a single line as below:
\noindent%
{
\begin{lstlisting}
type=SYSCALL msg=audit(1601405431.612391370:5893335): arch=40000028  per=800000 template=arducopter rep=1 stime=1601405431612391356 etime=1601405431612391367 ppid=1513 pid=1526 tid=1526 auid=1000 uid=0 gid=0 euid=0 suid=0 fsuid=0 egid=0 sgid=0 fsgid=0 tty=pts0 ses=1 comm="arducopter" exe="/home/pi/ardupilot/build/navio2/bin/arducopter" key=(null)\end{lstlisting}
}

Using the template and the compressed line of log, following three lines can be reconstructed. $\varnothing$ denotes values that could not be reconstructed and [min, max] enclose values for which range is known but not the exact value.

{
\begin{lstlisting}
type=SYSCALL msg=audit(1601405431.612391356:$\varnothing$): arch=40000028 syscall=4 per=800000 success=yes exit=8 a0=3 a1=$\varnothing$4 a2=1 a3=$\varnothing$ items=0 ppid=1513 pid=1526 tid=1526 auid=1000 uid=0 gid=0 euid=0 suid=0 fsuid=0 egid=0 sgid=0 fsgid=0 tty=pts0 ses=1 comm="arducopter" exe="/home/pi/ardupilot/build/navio2/bin/arducopter" key=(null)
type=SYSCALL msg=audit([1601405431.612391356, 1601405431.612391367]:$\varnothing$): arch=40000028 syscall=4 per=800000 success=yes exit=7 a0=4 a1=$\varnothing$ a2=1 a3=$\varnothing$ items=0 ppid=1513 pid=1526 tid=1526 auid=1000 uid=0 gid=0 euid=0 suid=0 fsuid=0 egid=0 sgid=0 fsgid=0 tty=pts0 ses=1 comm="arducopter" exe="/home/pi/ardupilot/build/navio2/bin/arducopter" key=(null)
type=SYSCALL msg=audit(1601405431.612391367:$\varnothing$): arch=40000028 syscall=4 per=800000 success=yes exit=7 a0=5 a1=$\varnothing$ a2=1 a3=$\varnothing$ items=0 ppid=1513 pid=1526 tid=1526 auid=1000 uid=0 gid=0 euid=0 suid=0 fsuid=0 egid=0 sgid=0 fsgid=0 tty=pts0 ses=1 comm="arducopter" exe="/home/pi/ardupilot/build/navio2/bin/arducopter" key=(null)\end{lstlisting}%
}%
As can be inferred from above, except an audit ID, all information can be reconstructed. Arguments that were not reconstructed here, were explicitly ignored in the template as they did not exist or were deemed irrelevant for forensic analysis. Event timings are inexact, but bounded. The range of uncertainty depends on whether \Sys or \Systwo is being used and the temporal policy.
Each process always has some unmodified audit entries, like exe from process spawn. Process wide constant entries like PROCTITLE can be reconstructed based on the audit information from the setup phase of the application.
The loss of exact event timings also loses the exact interleaving of events across different tasks.
But real-time tasks are designed to not have inter-task interference. Further, a successful iteration of the periodic task which meets its timing constraints has no further negative implications for future iterations.